\documentclass[twocolumn]{aastex63}

\usepackage{natbib}
\usepackage{hyperref}

\usepackage{booktabs}

\usepackage[T1]{fontenc}
\usepackage{graphicx}

\hypersetup{colorlinks=true, citecolor=blue, linkcolor=cyan, urlcolor=magenta}

\usepackage{epstopdf}

\newcommand{\h}{$^{\rm h}$}
\newcommand{\m}{$^{\rm m}$}
\newcommand{\s}{$^{\rm s}$}

\begin{document}

\title{The Star-Forming Interstellar Medium of Lyman Break Galaxy Analogs}

\author{John F. Wu}
\affiliation{Department of Physics and Astronomy, Rutgers, the State University of New Jersey, 136 Frelinghuysen Road, Piscataway, NJ 08854-8019, USA}
\affiliation{Center for Astrophysical Sciences, Department of Physics \& Astronomy, Johns Hopkins University, Baltimore, MD 21218, USA}
\email{jfwu@jhu.edu}

\author{Andrew J. Baker}
\affiliation{Department of Physics and Astronomy, Rutgers, the State University of New Jersey, 136 Frelinghuysen Road, Piscataway, NJ 08854-8019, USA}

\author{Timothy M. Heckman}
\affiliation{Center for Astrophysical Sciences, Department of Physics \& Astronomy, Johns Hopkins University, Baltimore, MD 21218, USA}

\author{Erin K. S. Hicks}
\affiliation{Department of Physics and Astronomy, University of Alaska, Anchorage, AK 99508-4664, USA}

\author{Dieter Lutz}
\affiliation{Max-Planck-Institut f\"ur extraterrestrische Physik, Giessenbachstrasse, D-85748 Garching, Germany}

\author{Linda J. Tacconi}
\affiliation{Max-Planck-Institut f\"ur extraterrestrische Physik, Giessenbachstrasse, D-85748 Garching, Germany}


\begin{abstract}
We present VLT/SINFONI near-infrared (NIR) integral field spectroscopy of six $z \sim 0.2$ Lyman break galaxy ``analogs'' (LBAs), from which we detect \ion{H}{1}, \ion{He}{1}, and [\ion{Fe}{2}] recombination lines, and multiple H$_2$ ro-vibrational lines in emission.
Pa$\alpha$ kinematics reveal high velocity dispersions and low rotational velocities relative to random motions ($\langle V_{\rm rot}/\sigma \rangle = 1.2 \pm 0.8$).
Matched-aperture comparisons of H$\beta$, H$\alpha$, and Pa$\alpha$ reveal that the nebular color excesses are lower relative to the continuum color excesses than is the case for typical local star-forming systems.
We compare observed \ion{He}{1}/\ion{H}{1} recombination line ratios to photoionization models to gauge the effective temperatures ($T_{\rm eff}$) of massive ionizing stars, finding the properties of at least one LBA are consistent with extra heating from an active galactic nucleus (AGN) and/or an overabundance of massive stars.
We use H$_2$~1-0~S$(\cdot)$ ro-vibrational spectra to determine rotational excitation temperature $T_{\rm ex} \sim 2000~\rm K$ for warm molecular gas, which we attribute to UV heating in dense photon-dominated regions.
Spatially resolved NIR line ratios favor excitation by massive, young stars, rather than supernovae or AGN feedback.
Our results suggest that the local analogs of Lyman break galaxies are primarily subject to strong feedback from recent star formation, with evidence for AGN and outflows in some cases. 
\end{abstract}

\section{Introduction} \label{sec:introduction}

Lyman break galaxies \cite[LBGs;][]{1996ApJ...462L..17S} are among the brightest and best-studied systems at high ($z \geq 3$) redshift. 
Efficient selection of LBGs has unveiled a wealth of information about the cosmic star formation rate (SFR) history \citep{2014ARA&A..52..415M} and imposed constraints on reionization \citep[e.g.,][]{2015ApJ...814...69A}. 
Despite this recent progress, a number of the interstellar medium (ISM) properties of LBGs remain poorly understood, as their small angular sizes \citep[see, e.g.,][]{2011A&A...532A..33G} and generally faint dust emission \citep[e.g.,][]{2015Natur.522..455C} are impediments to detailed study except in the rare cases of strongly lensed systems \citep[e.g.,][]{2004ApJ...604..125B,2007ApJ...665..936C} or exceptionally massive, resolved systems \citep[e.g.,][]{2012ApJ...758L...9M}. 
In particular, little is known from long-wavelength observations about the physical processes that drive evolution in LBGs.

For high-$z$ galaxies, stellar mass is assembled through in-situ star formation and accretion/merger activity \citep[e.g.,][]{2014ARA&A..52..291C,2014ARA&A..52..415M,2015ARA&A..53...51S,2017ApJ...837..150S}.
Interactions between galaxies can lead to enhanced star formation, an effect compounded by generally higher gas mass fractions and shorter free-fall times earlier in the Universe \cite[e.g.,][]{2007ARA&A..45..565M,2014PhR...539...49K,2018ApJ...861L..18U,2018MNRAS.477.2716K}.
The same gravitational instabilities that promote star formation can lead to an active galactic nucleus (AGN).
Both AGN feedback (driven by accretion onto the supermassive black hole) and star formation feedback (supernovae or stellar winds) can inject energy into and shock-heat the ISM gas \citep[see, e.g.,][]{2014ARA&A..52..589H,2016MNRAS.455.1830T,2017ApJ...841..101L,2018MNRAS.474.3673R}.
For high-redshift samples, Lyman and Balmer lines are shifted into optical and near-infrared (NIR) wavelengths, enabling the characterization of the ionized gas phase of their ISMs.
Observations of these lines have revealed strong outflows, mergers, and gas disks supported by rotation or turbulence/dispersion \citep{1996ApJ...462L..17S,2003ApJ...588...65S,2009ApJ...706.1364F,2015ApJ...799..209W,2018arXiv180809978W}.
Other phases of high-$z$ galaxies' ISMs are not as well understood, although recent instrumentation advances at millimeter wavelengths have allowed for more sensitive probes of the cool atomic and molecular gas in high-$z$ systems \citep[e.g.,][]{2013ARA&A..51..105C,2015ApJ...800...20G,2016ApJ...820...83S}.
Warm, potentially post-shock \textit{molecular} gas in LBGs has never been detected owing to the cosmological redshifting of the relevant spectral lines from rest-frame NIR to mid-infrared wavelengths.
Warm H$_2$ has been studied in detail for low-$z$ galaxy samples, but the lack of such data at high redshifts hinders our comprehension of how feedback processes affect different phases of the ISM at earlier epochs.

These limitations have prevented us from fully understanding the sources of ionization and excitation in the ISMs of high-redshift galaxies. 
Studies of star-forming LBGs imply that massive star binaries may contribute more ionizing emission at higher redshifts \citep[see, e.g.,][]{2013ApJ...774..152B,2016ApJ...826..159S}.
However, LBGs produce only modest amounts of energetic radiation through star formation in their disks \citep{2003ApJ...588...65S}, unlike the extreme merger-driven starbursts at $z \approx 2$ \citep[][]{2011ApJ...739L..40R}.
AGN are also known to be significant contributors to the total cosmic ionizing photon budget at $z \sim 2-4$ \cite[e.g.,][]{2018ApJ...863...63S}.
Simulations indicate that high-$z$ systems can grow via cold gas flows \citep{2005MNRAS.363....2K,2009MNRAS.395..160K,2009Natur.457..451D}, although accretion and merger-triggered shocks may also be important for understanding mass assembly in galaxies \citep{2003MNRAS.345..349B}. 
The nature of gas accretion and self-regulating feedback in distant galaxies can help explain the origins of their clumpy, high-pressure ISMs \citep[e.g.,][]{2010MNRAS.406..535P,2014MNRAS.443.3675M,2017MNRAS.464..491F,2019MNRAS.tmp.2058Z}.

The existence of analog populations in the local Universe gives us the opportunity to study rest-frame UV-selected galaxies in far greater detail than is generally possible at high redshift.
Previous works have identified low-$z$ systems that resemble high-$z$ galaxies in various regards,e.g., galaxies with clumpy morphologies \citep{2017MNRAS.464..491F} or high equivalent width Ly$\alpha$ emission \citep{2014ApJ...797...11O}).
Green pea galaxies \citep{2009MNRAS.399.1191C} and other systems selected on the basis of high nebular excitation \citep[e.g.,][]{2016ApJ...822...62B} exhibit high ionization parameter and electron density similar to those of distant star-forming galaxies, although their low masses ($\log(M_\star/M_\odot) \sim 9$) can hinder comparisons with the LBG population. 

From the Sloan Digital Sky Survey (SDSS) and \textit{Galaxy Evolution Explorer} (\textit{GALEX}) All-Sky Imaging Survey, \cite{2005ApJ...619L..35H} have identified a sample of $z < 0.3$  LBG analogs (LBGs) with unusually high FUV luminosities ($\log (L_{\rm FUV}/L_\sun) > 10.3$) and high FUV surface brightnesses.
This sample was later refined to include only the most ``supercompact'' UV-luminous galaxies with $\log (I_{\rm FUV}/ {L_{\sun}~\rm kpc^{-2}}) > 9$ \citep{2007ApJS..173..441H}. 
\cite{2009ApJ...706..203O} report follow-up observations of the supercompact LBA subsample with \textit{Hubble Space Telescope} (\textit{HST}) UV and optical imaging, revealing clumpy star formation and evidence for frequent mergers/interactions. 
Although morphological signatures of mergers are absent from the majority of high-$z$ LBGs, suggesting that rapid gas accretion must be fueling their high SFRs, \cite{2010ApJ...724.1373G} and \cite{2010ApJ...710..979O} show that LBAs lose their asymmetric merger-like features when artificially redshifted to $z \sim 2-4$.
Simulations of gas-rich mergers also reveal that interacting galaxies can masquerade as regularly rotating disks (with high velocity dispersions) when observed at lower spatial resolutions \citep{2016ApJ...816...99H}.
Without rest-UV observations sensitive to low surface brightnesses, it is difficult to conclude solely via morphology whether LBGs predominantly grow through merging activity or gas accretion.
However, resolved kinematics and studies of feedback from integral field spectroscopy can provide additional constraints on their evolution.

\cite{2010ApJ...724.1373G} investigate the Pa$\alpha$ (ionized gas) dynamics for a sample of 19 supercompact LBAs by using Keck/OSIRIS observations with adaptive optics (AO).
Their high angular resolution ($\sim 0.1\arcsec$) allows them to probe kinematics on $\sim 200$~pc scales without artificially smearing velocity structure over a large beam, but may also cause them to miss faint extended emission.
Velocity dispersions are very high, indicating that LBAs are dynamically young systems characterized by strong starbursts and merger activity.
By simulating observations of these LBAs at $z = 2.2$, the authors find strong morphological resemblances to high-redshift LBGs, and show that they can resolve complex kinematic structures that would otherwise be smoothed out at lower physical resolutions.

\begin{deluxetable*}{l rr r R R R R}
	\tablewidth{0pt}
	\tablecolumns{8}
	\tablecaption{VLT/SINFONI observations \label{tab:vlt observations}}
	\colnumbers
	\tablehead{
		\colhead{Object}  &
		\colhead{$\alpha_{\rm J2000}$} &
		\colhead{$\delta_{\rm J2000}$} &
		\colhead{$z$} &
		\colhead{$\log (M_*/M_\sun)$} &
		\colhead{12+log(O/H)} &
		\colhead{log(SFR/$M_\sun~\rm yr^{-1}$)} &
		\colhead{log(sSFR/Gyr$^{-1}$)}
	}
	\startdata
	015028 (J0150) & 01\h 50\m 28\fs 39 &  13\arcdeg 08\arcmin 58\farcs 4 & 0.147 & 10.3 &  8.39   &  1.71 &  0.41 \\
	021225 (J0212) & 02\h 12\m 25\fs 82 & -08\arcdeg 01\arcmin 22\farcs 8 & 0.114 &  9.9 &  8.44   &  1.15 &  0.25 \\
	021348 (J0213) & 02\h 13\m 48\fs 53 &  12\arcdeg 59\arcmin 51\farcs 5 & 0.219 & 10.5 &  8.74   &  1.55 &  0.05 \\
	143417 (J1434) & 14\h 34\m 17\fs 15 &  02\arcdeg 07\arcmin 42\farcs 5 & 0.180 & 10.7 &  8.65   &  1.30 & -0.40 \\
	210358 (J2104) & 21\h 03\m 58\fs 74 &  07\arcdeg 28\arcmin 02\farcs 4 & 0.137 & 10.9 &  8.70   &  2.03 &  0.14 \\
	211531 (J2116) & 21\h 15\m 31\fs 01 & -07\arcdeg 27\arcmin 39\farcs 1 & 0.138 & 10.7 & $\cdots$&  1.26 & -0.44 \\
	\enddata
	\tablecomments{Ancillary data from \cite{2005ApJ...619L..35H} and \cite{2009ApJ...706..203O}. 
		Columns are (1) object IDs, (2) Right Ascension, (3) Declination, (4) spectroscopic redshift, (5) stellar mass, (6) nebular-phase metallicity, (7) SFR from H$\alpha$ and 24~\micron{} flux, and (8) specific SFR $\equiv$ SFR/$M_*$. 
		The median SFR for our sample is $28~M_\sun~\rm yr^{-1}$, and the median sSFR is $1.2\rm~Gyr^{-1}$.
		}
\end{deluxetable*}

Low-$z$ LBAs resemble the LBG population in many other ways; the populations have similar UV-optical colors, dust extinctions, metallicities, SFRs, physical half-light radii, and emission-line velocity widths.
\cite{2007ApJS..173..457B} investigate the radio continuum properties of supercompact UV-selected LBAs using Very Large Array (VLA) observations, finding that they are characterized by low dust attenuations and a depressed radio to UV ratio. 
Using CO$(1-0)$ spectral line observations from the Combined Array for Research in Millimeter-wave Astronomy (CARMA), \cite{2014MNRAS.442.1429G} find that LBAs adhere to the low-redshift Kennicutt-Schmidt \citep{1998ApJ...498..541K} law and consume their gas on short ($0.1-1$~Gyr) timescales, while harboring large ($0.2-0.6$) gas mass fractions.
\cite{Contursi+17} use \textit{Herschel} PACS and IRAM Plateau de Bure Interferometer (PdBI) observations to provide additional evidence that physical conditions in the ISMs of LBAs are more similar to those of high-redshift sources than local star-forming galaxies.

We have selected for follow-up six LBAs drawn from the original \cite{2005ApJ...619L..35H} sample, of which four satisfy the \cite{2007ApJS..173..441H} supercompact definition (and thus also have \textit{HST} imaging available). 
These six LBAs had been targeted for earlier, long-wavelength follow-up observations on the basis of their rest-UV luminosities and colors (or equivalently, attenuations), which predicted that their far-IR luminosities would be among the highest in the parent sample \citep[e.g.,][]{1999ApJ...521...64M}, and indeed all but one\footnote{021225 = J0212} yielded (low-S/N) detections when {\it IRAS} $60\,{\rm \mu m}$ scans were reprocessed using the SCANPI tool \citep{2009ASPC..411...33A}. 
\cite{Contursi+17} later confirmed their high dust-obscured SFRs using \textit{Herschel} observations.
We have obtained VLT/SINFONI integral field spectroscopy at NIR wavelengths for all six targets, providing spatially resolved views of hydrogen and helium recombination in addition to H$_2$ ro-vibrational lines.
In Table~\ref{tab:vlt observations}, we list some of our sample's properties while the new observations and data reduction are described in Section~\ref{sec:observations}.

This work aims to explore some of the rest-frame NIR properties of LBGs and their analogs that are difficult or impossible to determine at higher redshifts.
In Section~\ref{sec:kinematics}, we assess ionized gas dynamics using the bright Pa$\alpha$ line.
In Section~\ref{sec:results}, we use multi-component nebular lines to characterize the global dust attenuations and effective temperatures of star-forming clumps in the ISM.
Additionally, we report results on the \textit{warm} molecular ISM phase for the first time in LBAs.
In Section~\ref{sec:H2 masses}, we compare the cool and warm molecular gas phases of the ISMs of LBAs, and in Section~\ref{sec:line-ratios}, we investigate spatially resolved differences in line flux ratios.
In Section~\ref{sec:LBGs-LBAs}, we discuss similarities and differences between LBGs and their low-$z$ analogs, and we summarize our conclusions in Section~\ref{sec:conclusions}.
Throughout the paper, we assume a concordance $\Lambda$CDM cosmology ($H_0 = 70~\rm km~s~Mpc^{-1}$, $\Omega_M=0.3$, and $\Omega_\Lambda = 0.7$), such that $1\arcsec $ corresponds to a projected distance of $3.3~$kpc at $z = 0.2$.
All magnitudes are Vega-relative.

\section{Observations and data reduction} \label{sec:observations}

\subsection{VLT/SINFONI observations}

VLT/SINFONI observations of our targets were obtained in service mode
between 2008 June 16 and 2008 August 11 via ESO programme 081.B-0947
(PI: Baker).
We used SINFONI \citep{2003SPIE.4841.1548E,2004Msngr.117...17B} in its seeing-limited mode, with a $0.25\arcsec~{\rm pixel^{-1}}$ image scale delivering an $8\arcsec \times 8\arcsec$ field of view.
(Most of our targets were sufficiently compact to lie well within this field of view; for J2116, which comprises a pair of objects, our pointing center lay closer to J2116a, with the result that J2116b ended up falling partly outside the field.)  
For the redshifts of our sample, observing the Pa$\alpha$ line required observations in $K$ band.  
Each observation began with the acquisition of a bright ($11.8 \leq K \leq 15.4$), nearby star, which was also used as a PSF reference, before the telescope was slewed to the science target.  
Using the \texttt{SINFONI\_ifs\_obs\_GenericOffset} observing template, we then proceeded to take six 600\,s exposures switched between object (O) and sky (S) positions in an O-S-O-O-S-O sequence; offsets to sky positions were $\sim 20~{\rm arcsec}$, and additional sub-arcsecond dithers between exposures were introduced to facilitate data processing.  
The choice of a 600\,s exposure time was motivated by our interest in enabling reliable background subtraction while remaining in the background-limited regime for $K$ band.  
LBAs J0212, J0213, and J2104 were observed once apiece, i.e., giving 2400\,s on-source; J0150 and J2116 were observed twice apiece, giving 4800\,s on-source; and J1434 was observed 1.5 times, giving 3600\,s on-source.
Weather conditions during the observations were generally good, with clear skies and $K$-band seeing of $0.8\arcsec$ or better.

\subsection{Data reduction} \label{sec:data-reduction}

Data reduction was completed with the custom software package SPRED \citep{2006NewAR..50..398A}.  
SPRED performs the typical data reduction steps for NIR spectra with additional routines necessary to reconstruct the data cube.  
The final pixel scale is 0.125\arcsec{} for all data cubes.
We measure the PSF by fitting an elliptical Gaussian to the PSF reference source, which has been averaged over all channels. 
The PSF is typically characterized by FWHM~$\sim 0.5-0.7\arcsec$ for our observations.
In addition, the residuals from night sky emission lines are minimized using the methods outlined in \cite{2007MNRAS.375.1099D} through careful wavelength registration and background subtraction, which exploits the sky frames interspersed with on-source exposures.  
Telluric correction and flux calibration were performed using A- and B-type stars, with flux calibration estimated to be accurate to within 10\%.

\section{Ionized gas kinematics} \label{sec:kinematics}

\begin{figure*}
	\plotone{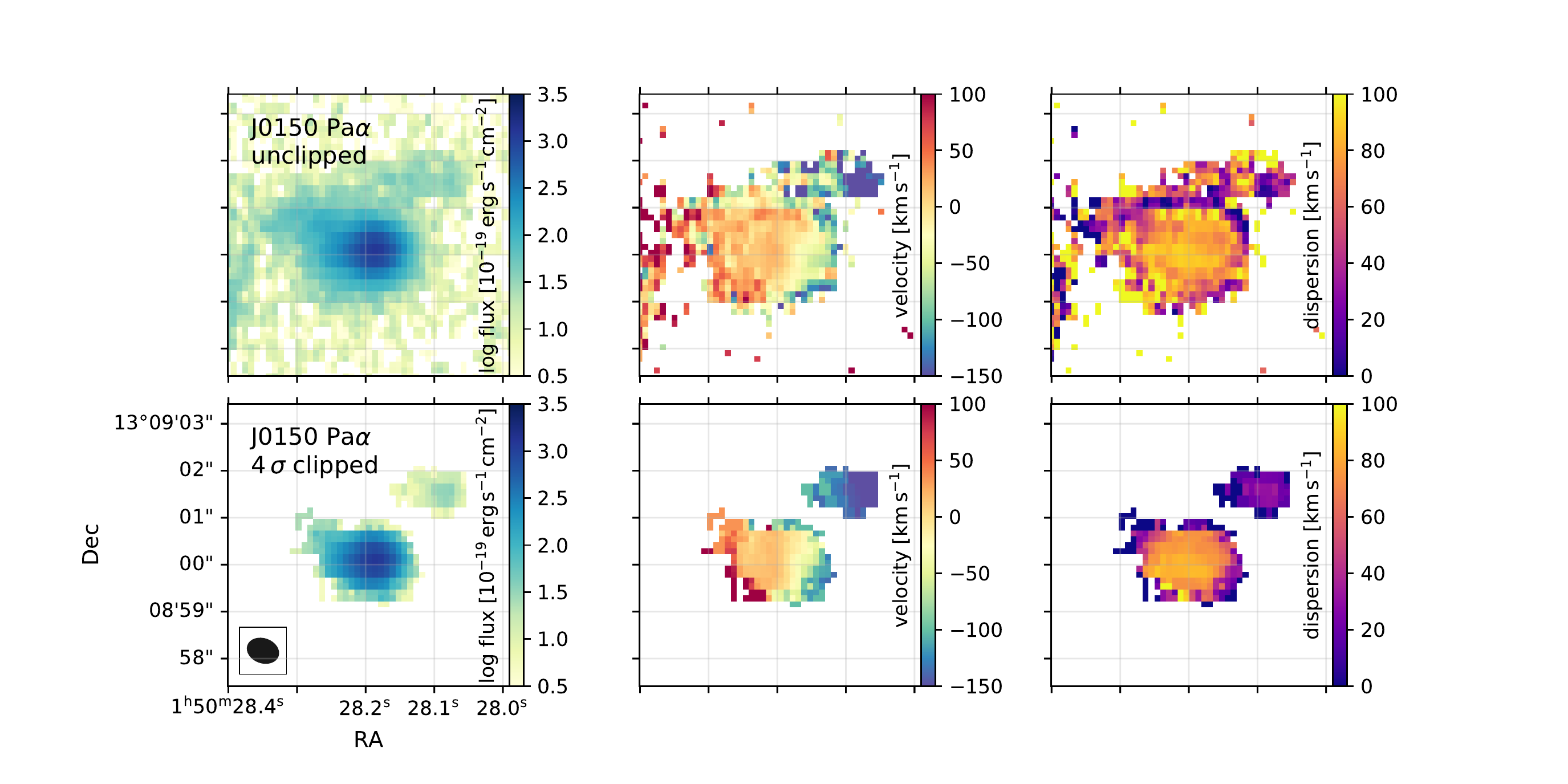}
	\plotone{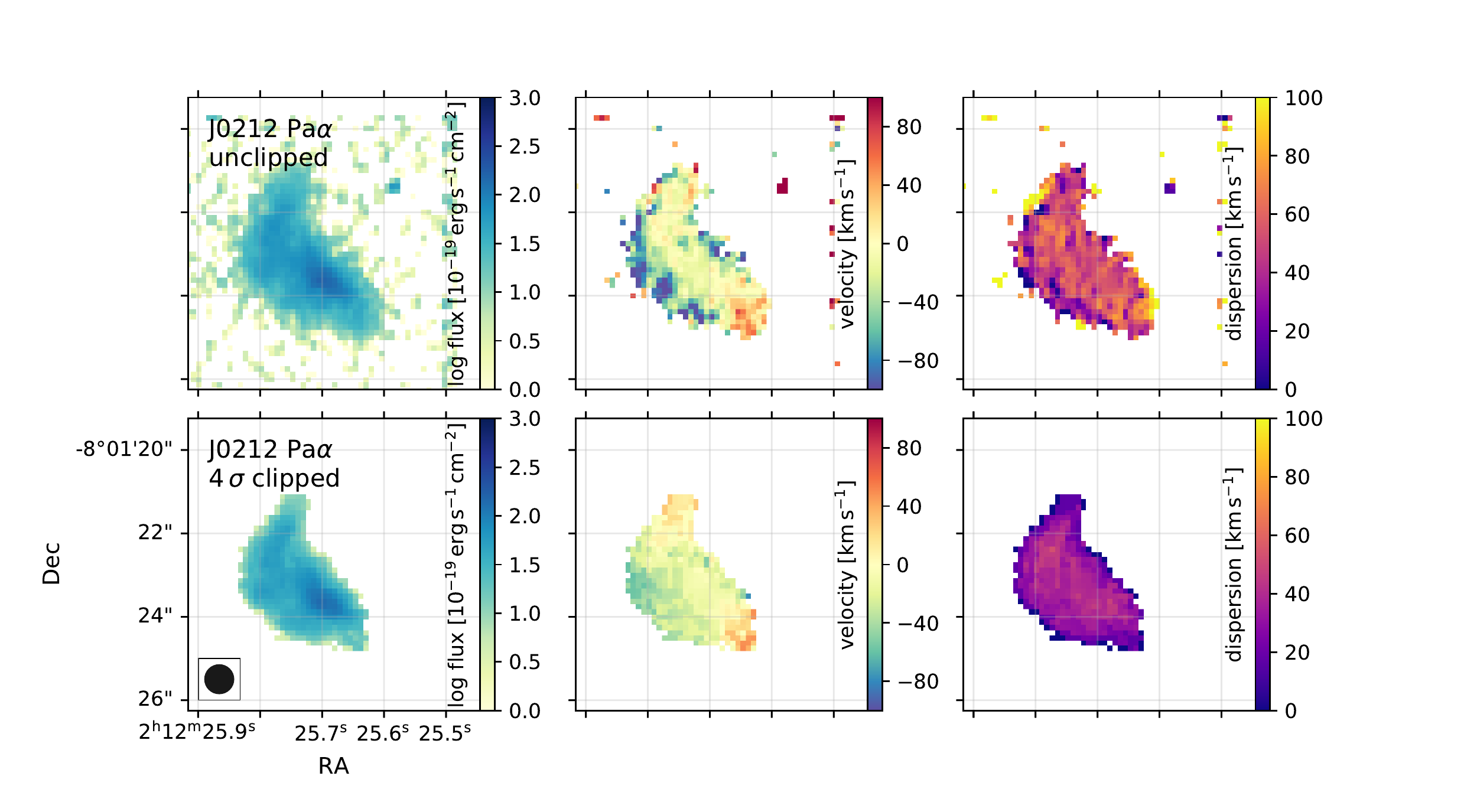}
	\plotone{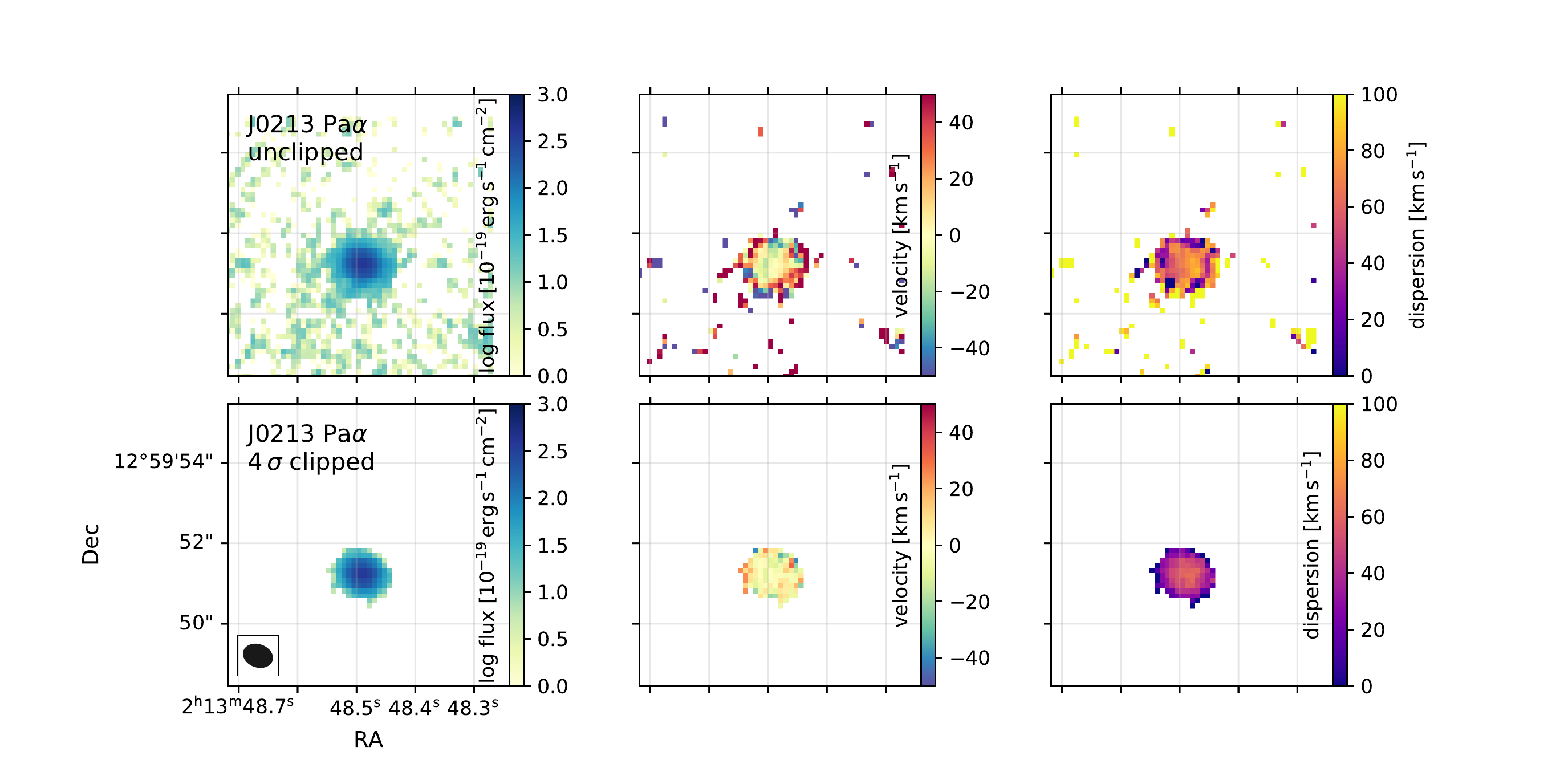}
\end{figure*}
\begin{figure*}
	\plotone{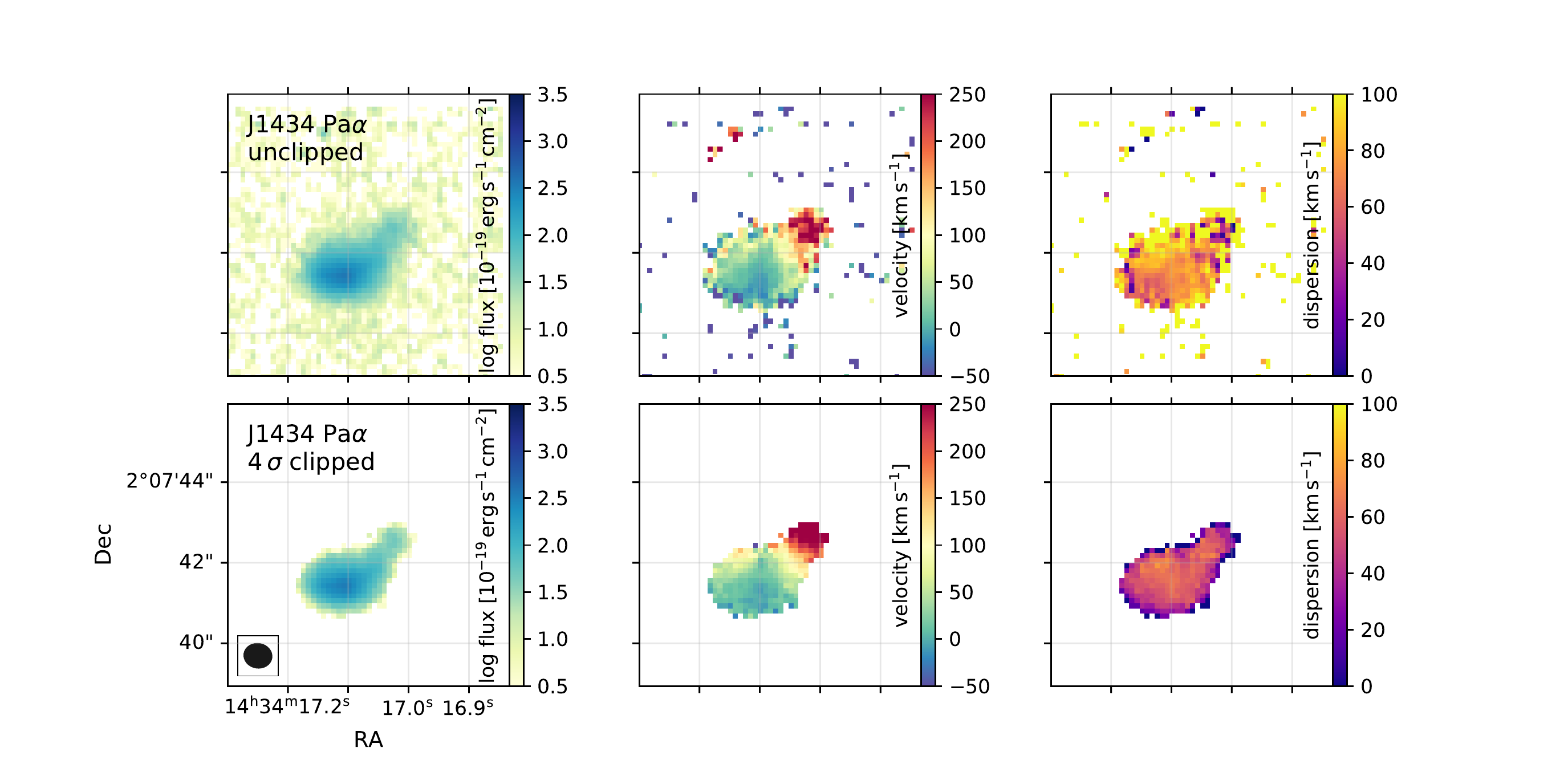}
	\plotone{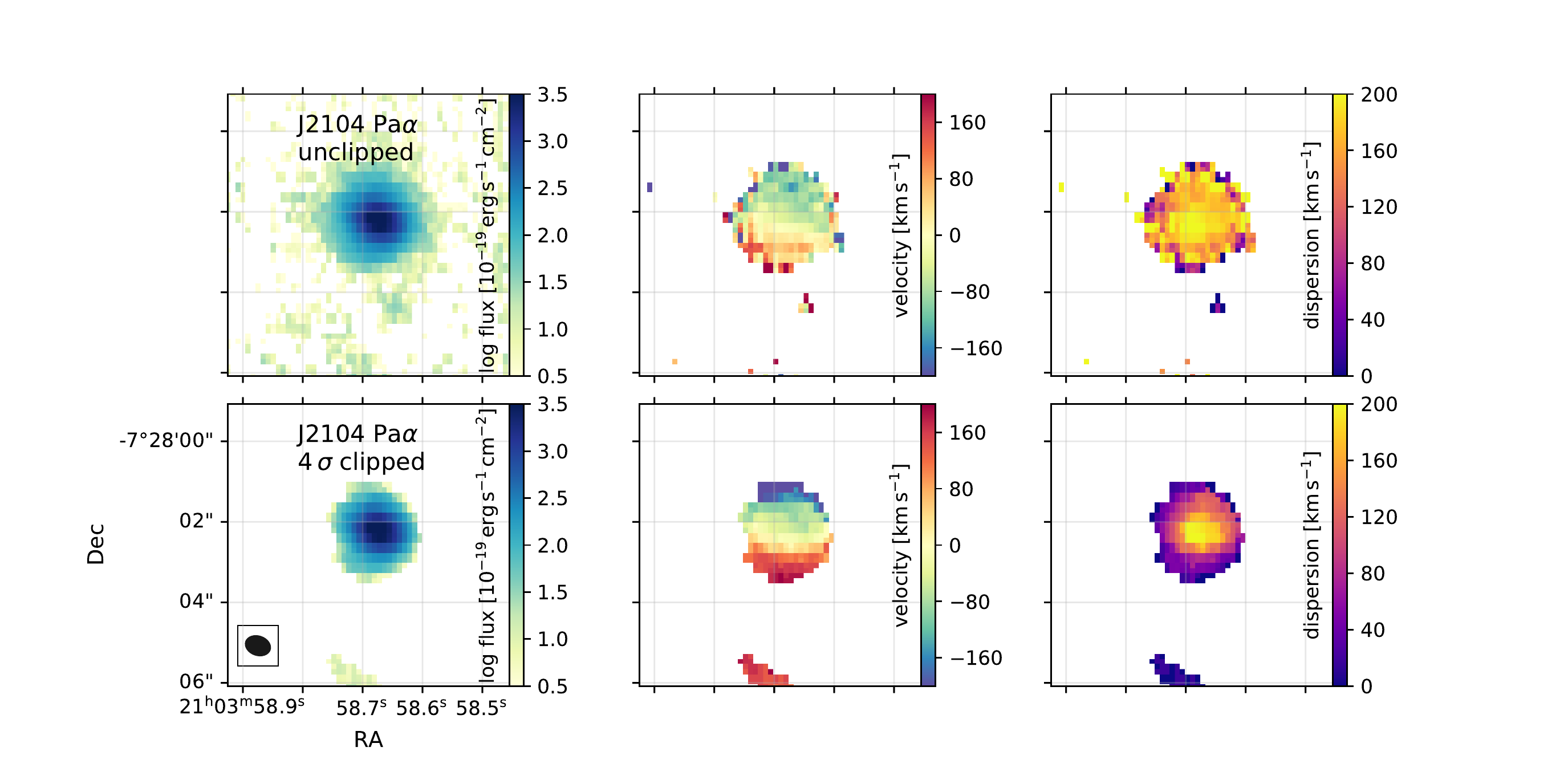}
	\plotone{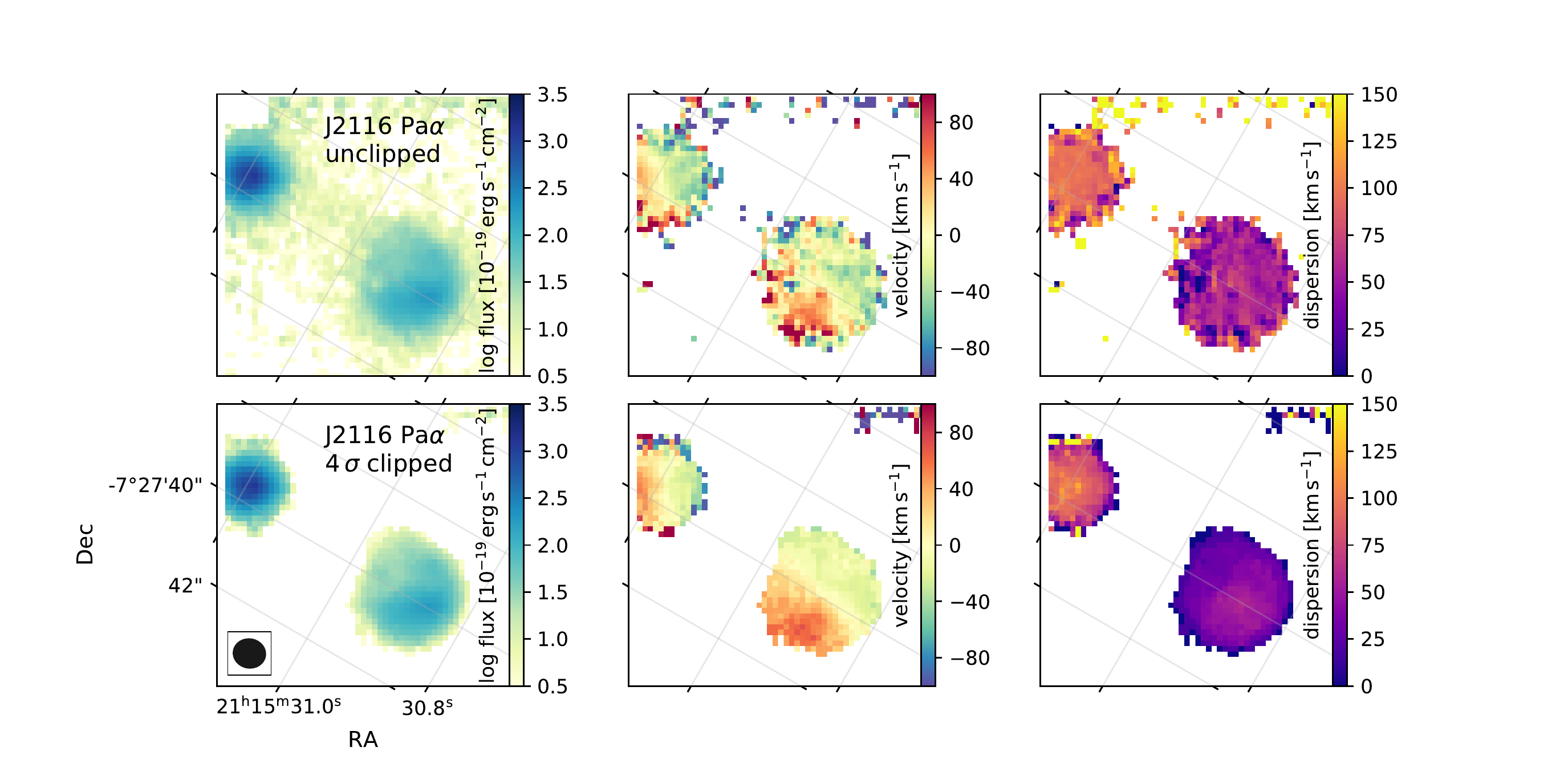}
	\caption{\label{fig:moment-maps}
		In each set of panels, we show VLT/SINFONI Pa$\alpha$ integrated flux maps (\textit{left}), velocity maps (\textit{center}), and dispersion maps (\textit{right}), generated without (upper) and with (lower) clipping of pixels below $4~\sigma_{\rm rms,~chan}$.
		Unclipped velocity and dispersion maps have been masked at the $2~\sigma$ integrated flux level only for visualization purposes.
		The PSF is shown at the lower left as a black ellipse.
	}
\end{figure*}

We investigate ionized gas kinematics for our sample of seven resolved LBAs.
We produce moment maps of the Pa$\alpha$ integrated flux, rotation velocity ($V_{\rm rot}$), and velocity dispersion ($\sigma$) for each system (Figure~\ref{fig:moment-maps}).
We also use a disk modeling program to measure ordered rotation and velocity dispersions, and find general agreement with the moment map analysis.
Velocity dispersions are high, $\sigma \approx 50-110 {\rm ~km~s^{-1}}$, and ordered rotation-to-dispersion velocity ratios are in line with similar LBA samples in the literature: $\langle v/\sigma \rangle \approx 1.2 \pm 0.8$.
LBA kinematic properties are summarized in Table~\ref{tab:kinematics}.
In the subsections below, we describe our analyses and comparisons to previous results.

\subsection{Moment maps} \label{sec:moment-maps}

We generate moment maps from median-subtracted VLT/SINFONI spectral cubes centered on the Pa$\alpha$ line. 
In the upper panels of Figure~\ref{fig:moment-maps}, we show the integrated flux, velocity, and dispersion maps (ordered left to right).
The $4~\sigma$ Pa$\alpha$ surface brightness detection limit ranges between $(0.8 - 2.5)\times 10^{-18} {~\rm erg~s^{-1}~cm^{-2}}$ per pixel.
We produce a separate set of \textit{clipped} moment maps in which low signal-to-noise ratio (SNR) pixels are masked;\footnote{We estimate the rms noise, $\sigma_{\rm rms}$ by removing $3\,\sigma$ outliers with a median absolute deviation (MAD) cut and taking a standard deviation for each channel; outliers are only removed for the purpose of estimating $\sigma_{\rm rms}$.} 
clipped moment maps are shown in the lower panels of Figure~\ref{fig:moment-maps}.

Unclipped and clipped moment maps are characterized by several visual differences.
The unclipped moment maps trace extended low-surface brightness emission in a few cases, but are generally noisier than their clipped counterparts.
Clipped rotation velocity maps have smoother velocity fields, and dispersion maps are peaked towards the centers.
In some cases, the clipped velocity maps may exclude \textit{bona fide} low-SNR emission at extremal velocities, which can have the effect of diminishing the rotation velocity gradient.

We compute the ratio of ordered rotation velocity to velocity dispersion, $V_{\rm rot}/\sigma$, where $V_{\rm rot} \equiv (v_{\rm max} - v_{\rm min})/2$ is half the range in the flux-weighted velocity field,\footnote{$v_{\rm max}$ and $v_{\rm min}$ are the maximum and minimum velocities in the first moment map, respectively.} and $\sigma$ is the Pa$\alpha$ flux-weighted velocity dispersion averaged over all pixels.
For ordinary star-forming galaxies, $V_{\rm rot}/\sigma$ measures the degree of rotational support.
Strong turbulent processes, such as the injection of energy from star formation or AGN feedback, will physically puff up galaxy disks and inflate $\sigma$.
For interacting systems, $V_{\rm rot}/\sigma$ can skew low because $\sigma$ is elevated by disks with multiple inclinations or even counterrotation, relative to a single gas disk.
Such galaxy interactions can also promote star formation and escalate feedback processes.

In Table~\ref{tab:kinematics}, we list each LBA and its ordered velocity, dispersion, and $V_{\rm rot}/\sigma$ ratio.
After applying an averaged inclination correction factor of $4/\pi \approx 1.3$ \cite[see, e.g., the Appendix of][]{2009ApJ...697.2057L}, we measure clipped $\langle V_{\rm rot} \rangle = 160 \pm 99 {\rm ~km~s^{-1}}$ and unclipped $\langle V_{\rm rot} \rangle  = 145 \pm 54 {\rm ~km~s^{-1}}$.
Our sample is characterized by high gas dispersions: clipped and unclipped dispersions are $\langle \sigma \rangle = 76 \pm 45 {\rm ~km~s^{-1}}$ and $89 \pm 42{\rm ~km~s^{-1}}$ respectively, which greatly exceed the $\sim 5-15 {\rm ~km~s^{-1}}$ dispersions characteristic of low-$z$ disk galaxies \citep{2006ApJ...638..797D}.
We find $\langle V_{\rm rot}/\sigma \rangle = 2.2 \pm 0.9$ from clipped moment maps and $1.8 \pm 0.7$ from unclipped moment maps.

\begin{deluxetable*}{l rrr c rrr c rrr}
	\tablewidth{0pt}
	\tablecolumns{12}
	\setlength{\tabcolsep}{9pt}
	\tablecaption{Ordered velocities and dispersions \label{tab:kinematics}}
	\tablehead{
		\colhead{} &
		\multicolumn{3}{c}{Clipped} & &
		\multicolumn{3}{c}{Unclipped}  & & 
		\multicolumn{3}{c}{Modeled} \\
		\cline{2-4} 
		\cline{6-8}
		\cline{10-12}
		\colhead{LBA} & 
		\colhead{$V_{\rm rot}$} &
		\colhead{$\sigma$} &
		\colhead{$V_{\rm rot}/\sigma$} & &
		\colhead{$V_{\rm rot}$} &
		\colhead{$\sigma$} &
		\colhead{$V_{\rm rot}/\sigma$} & &
		\colhead{$V_{\rm rot}$} &
		\colhead{$\sigma$} &
		\colhead{$V_{\rm rot}/\sigma$} \vspace{-0.75em}\\
		\colhead{} &
		\colhead{$\rm [km~s^{-1}]$} &
		\colhead{$\rm [km~s^{-1}]$} &
		\colhead{} & &
		\colhead{$\rm [km~s^{-1}]$} &
		\colhead{$\rm [km~s^{-1}]$} &
		\colhead{} & &
		\colhead{$\rm [km~s^{-1}]$} &
		\colhead{$\rm [km~s^{-1}]$} &
		\colhead{}
	}
	\decimals 
	\startdata
	J0150 & 161.3 &  75.2 & 2.14 && 61.7 & 80.5 & 0.77 && $67.7 \pm 1.3$ & $100.7 \pm 0.4$ & $0.67 \pm 0.01$\\
	J0212 &  63.6 &  36.0 & 1.77 && 86.3 & 51.4 & 1.68 && $33.0 \pm 1.7$ & $49.6 \pm 0.6$ & $0.67 \pm 0.04$ \\
	J0213 &  35.4 &  49.9 & 0.71 && 72.4 & 64.2 & 1.13 && $ 8.5 \pm 7.7$ & $78.4 \pm 3.2$ & $0.11 \pm 0.10$ \\
	J1434\tablenotemark{\dag} &  182.5 & 56.5 & 3.23 && 135.5 & 73.9 & 1.83 && $100.1 \pm 0.1$ & $87.4 \pm 0.4$ & $1.15 \pm 0.01$ \\
	J2104 & 251.4 & 177.9 & 1.41 && 192.2 & 183.0 & 1.05 && $258.1 \pm 2.8$ & $98.7 \pm 1.0$ & $2.62 \pm 0.04$ \\
	J2116a &  56.8 &  43.1 & 1.32 && 138.0 & 58.0 & 2.38 && $108.2 \pm 5.8$ & $47.6 \pm 0.6$ & $2.27 \pm 0.12$ \\
	J2116b & 128.5 &  94.3 & 1.36 && 112.3 & 108.5 & 1.03 && $107.2 \pm 2.2$ & $112.6 \pm 2.2$ & $0.95 \pm 0.03$\\
	\enddata
	\tablecomments{
		Kinematics measured from $4~\sigma$ clipped and unclipped Pa$\alpha$ moment maps (uncorrected for inclination).
		Image segmentation from the \texttt{photutils} package has been applied to isolate emission from individual components. 
		Kinematics are also inferred from GalPaK3D modeling, where inclination, PSF, and instrumental broadening effects have been taken into account.
	}
	\tablenotetext{\dag}{
		Kinematic information shown for the full blended system. 
		For only the main central component, isolated using a manual mask applied to the clipped moment maps, we find $v = 99.5~{\rm km~s^{-1}}$, $\sigma = 57.2~{\rm km~s^{-1}}$, and $v/\sigma = 1.74$.
	}
\end{deluxetable*}

\subsection{Disk modeling} \label{sec:disk-modeling}

Kinematics measured from moment maps may be unreliable due to beam smearing effects, which can artificially transform rotational kinematics into dispersive motions \citep[see, e.g.,][]{2011ApJ...741...69D}.\footnote{We repeat our analysis with the central portions of each system masked, and find that the reduction in $\sigma$ is less than 10\%. This implies that beam smearing effects on the moment maps are small, and we note that disk modeling is able to account for other instrumental and systematic effects as well.}
Merging galaxies can appear as disks, and any misidentified mergers could lead to an artificial increase in $\sigma$ \citep[e.g.,][]{2010ApJ...724.1373G,2016ApJ...816...99H}.
These concerns are warranted in particular at higher redshifts, where the angular sizes of distant sources become comparable to the observational resolution.
In such cases, it is often advantageous to use sophisticated models that can factor in the PSF and instrument's spectral broadening while constraining the disk's intrinsic kinematics \cite[e.g.,][]{2009ApJ...697..115C,2012ApJ...758..106K,2013ApJ...767..104N,2014ApJ...785...75G,2015MNRAS.452..986S,2016ApJ...830...14S,2017ApJ...843...46S,2018MNRAS.474.5076J}.

We use the \texttt{GalPaK3D} software \citep{2015AJ....150...92B} to model the kinematics of our LBA Pa$\alpha$ line emission directly from the SINFONI data cubes.
\texttt{GalPaK3D} can account for seeing effects and instrumental line broadening, which allows us to extract not only kinematic quantities ($V_{\rm rot}$, $\sigma$, and velocity turnover radius), but also geometric and shape parameters (inclination, position angle, and half-light radius).
We use the default exponential surface brightness model with an arctan rotation curve and thick disk \citep[see, e.g.,][]{2008gady.book.....B,2008ApJ...687...59G}.

We fit median-subtracted Pa$\alpha$ cubes with kinematic models using default uniform priors. 
If the fits are poor or if the Markov Chain Monte Carlo (MCMC) routine does not converge after 4000 steps, we impose additional constraints, such as priors on the centroid positions or velocity ranges (inferred from moment maps) or masking out companion sources.
We estimate the mean and standard deviation for $V_{\rm rot}$ and $\sigma$ using the final 700 steps of the converged MCMC chain.
For a few LBAs in our sample, the derived half-light radii are comparable to the seeing ($\sim 0.7\arcsec$), which may bias the inclination-dependent $V_{\rm rot}$ high \citep[][]{2015AJ....150...92B}.

\texttt{GalPaK3D} model results for our sample are shown in Table~\ref{tab:kinematics}.
Velocity dispersions range from about $50 - 110 ~{\rm km~s^{-1}}$, which agree with inclination-corrected dispersions measured from moment maps.
We find an unweighted average $\langle \sigma \rangle = 82 \pm 23 {\rm ~km~s^{-1}}$.
Modeled kinematics also yield $V_{\rm rot}/\sigma$ consistent with those from our previous analysis; $\langle V_{\rm rot}/\sigma \rangle = 1.2 \pm 0.8$ confirms that, on average, the LBAs in our sample are only marginally supported by rotation.

\subsection{Comparison to previous results}

Our analysis builds on the results of \cite{2009ApJ...699L.118B} and \cite{2010ApJ...724.1373G}, who have observed 19 LBAs with Keck/OSIRIS using adaptive optics (AO).
Four members of our sample also belong to theirs, although we note that our selection criteria tend to favor systems with higher dust content and lower UV surface brightnesses, so the two samples' average properties differ slightly.
Our SINFONI observations have coarser resolution (in addition to larger pixel scales and wider FOVs), and our integration times are generally longer (e.g., $2400-4800$\,s compared to $1500-2700$\,s for \citealt{2010ApJ...724.1373G}).
Our Pa$\alpha$ detection thresholds correspond to $6~\sigma$ SFR surface densities of $\sim 0.01~M_\sun~{\rm yr}^{-1}$ (assuming Case B recombination and a \citealt{2003PASP..115..763C} IMF; \citealt{2012ARA&A..50..531K}), nearly an order of magnitude deeper than the \cite{2010ApJ...724.1373G} AO-assisted observations.
We find evidence for extended or clumpy low-surface brightness emission that is missing in the shallower observations (e.g., in J0150 and J2104).
Since we are able to detect faint emission at larger radii and rotation velocities, we also measure higher $V_{\rm rot}$ \citep[while $\sigma$ does not change;][]{2009ApJ...699L.118B,2010ApJ...724.1373G}.

Other works have explored the kinematic properties traced by ionized gas in high-$z$ star-forming systems.
The SINS survey targeted massive $z \sim 2$ galaxies that are physically larger and have higher rotation velocities than typical LBGs/LBAs \cite[$V_{\rm rot} \sim 150 - 300 {\rm ~km~s^{-1}}$;][]{2009ApJ...697..115C}.
\cite{2009ApJ...706.1364F} report $\langle V_{\rm rot}/\sigma \rangle \sim 4.4$ for the entire sample, and also note that approximately a third are dispersion-dominated ($V_{\rm rot}/\sigma < 1$).
The KMOS$^{\rm 3D}$ program \citep{2015ApJ...799..209W}, which surveyed a variety of $0.7< z < 2.7$ objects, also found mostly rotation-supported galaxies (e.g., $\langle V_{\rm rot}/\sigma \rangle \sim 5$ at $z \sim 1$ and $\langle V_{\rm rot}/\sigma \rangle \sim 3$ at $z \sim 2$).
The MOSDEF survey of $z \sim 1-3$ galaxies includes systems with diverse dust properties, SFRs, and masses \citep{2015ApJS..218...15K}; these galaxies are characterized by median $V_{\rm rot}/\sigma \sim 2$ \citep[e.g.,][]{2016ApJ...819...80P}.
For galaxies at given stellar mass, velocity dispersions increase with redshift, a trend that is also consistent with the observed increase in SFR with $\sigma$ \citep[see also, e.g.,][]{2010ApJ...724.1373G,2014MNRAS.437.1070G}.

Our sample of low-$z$ LBAs have similar kinematic properties to $z \sim 2$ MOSDEF systems at the same $M_{\star}$ and sSFR; however, high-$z$ massive ($\log(M_{\star}/M_{\sun}) \sim 11$) galaxies tend to have more ordered velocity fields, suggestive of disk settling \citep[e.g.,][]{2015ApJ...799..209W,2019arXiv190209554P}.
LBAs also have properties similar to those of $z \sim 2-3$ galaxies selected via a generalized Lyman-break technique \citep[e.g., morphologies, physical sizes, and stellar masses;][]{2009ApJ...697.2057L}.
The authors find $\sigma \sim 60-100~\rm km~s^{-1}$ (similar to our sample), and low velocity-to-dispersion ratio (i.e., inclination corrected $\langle v/\sigma \rangle \sim 0.3$).
However, the surface brightnesses probed by \cite{2009ApJ...697.2057L} are $\sim 30\times$ shallower than in our observations, which may explain why their sample appears to lack the rotational kinematic support present in our sample of LBAs.

In summary, our conclusions are consistent with those of other LBA studies (i.e., \citealt{2009ApJ...699L.118B,2010ApJ...724.1373G}).
LBAs and $z \sim 2$ samples of $\log(M_\star /M_\sun) \sim 10.5$ galaxies \citep[e.g.,][]{2009ApJ...697.2057L,2016ApJ...819...80P} have qualitatively similar kinematic properties, although higher-mass systems tend to be characterized by stronger rotational support \citep[$V_{\rm rot}/\sigma \gtrsim 3$;][]{2009ApJ...706.1364F,2015ApJ...799..209W}.
LBAs are marginally rotation dominated, similar to other galaxy samples found on the $z \sim 1-2$ SFR$-M_{\star}$ sequence \citep[e.g.,][]{2015ApJ...799..209W,Contursi+17}, such that their kinematics and star formation properties truly resemble those of high-$z$ systems.

\section{The multiphase ISM} \label{sec:results}

\begin{figure*}[ht!]
	\plotone{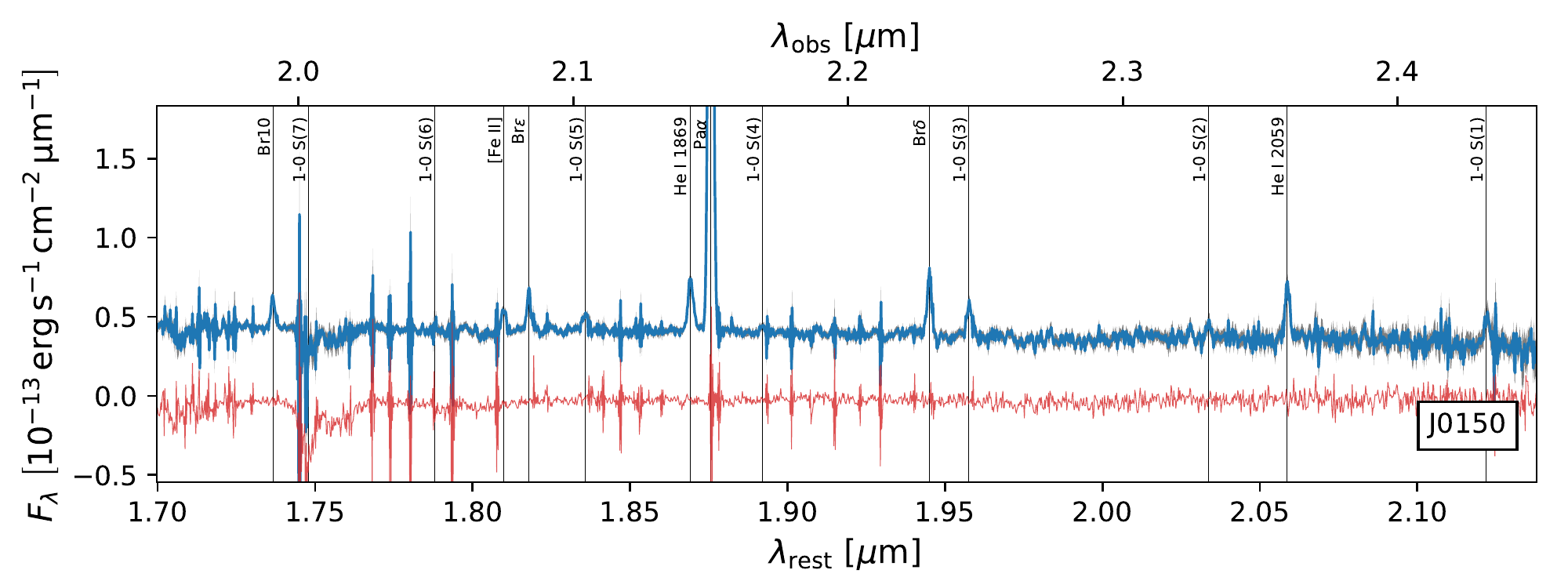}	
	\plotone{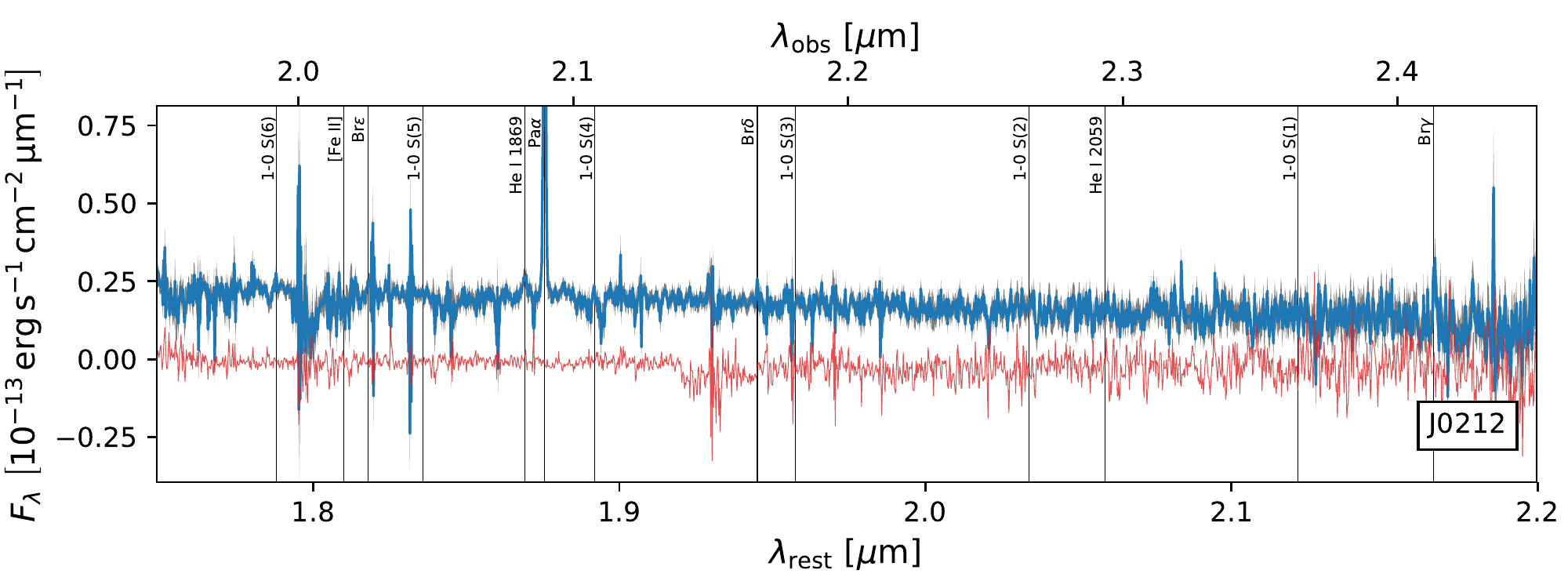}
	\plotone{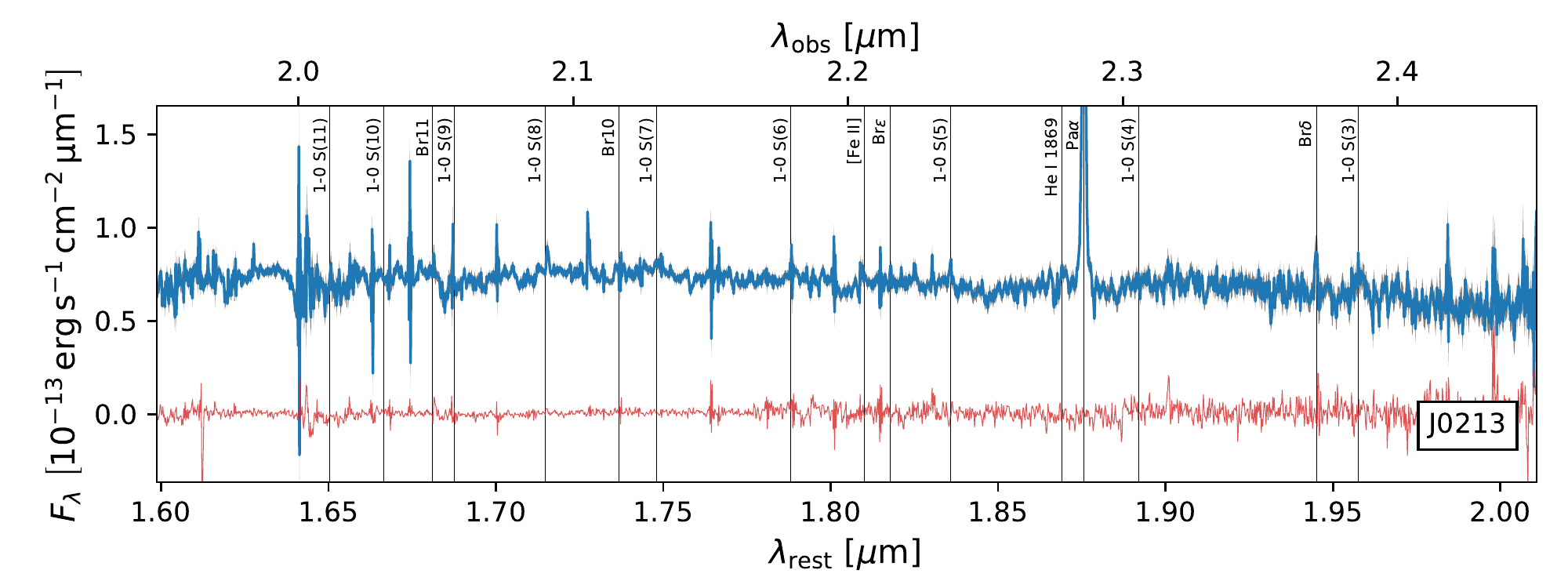}
	\plotone{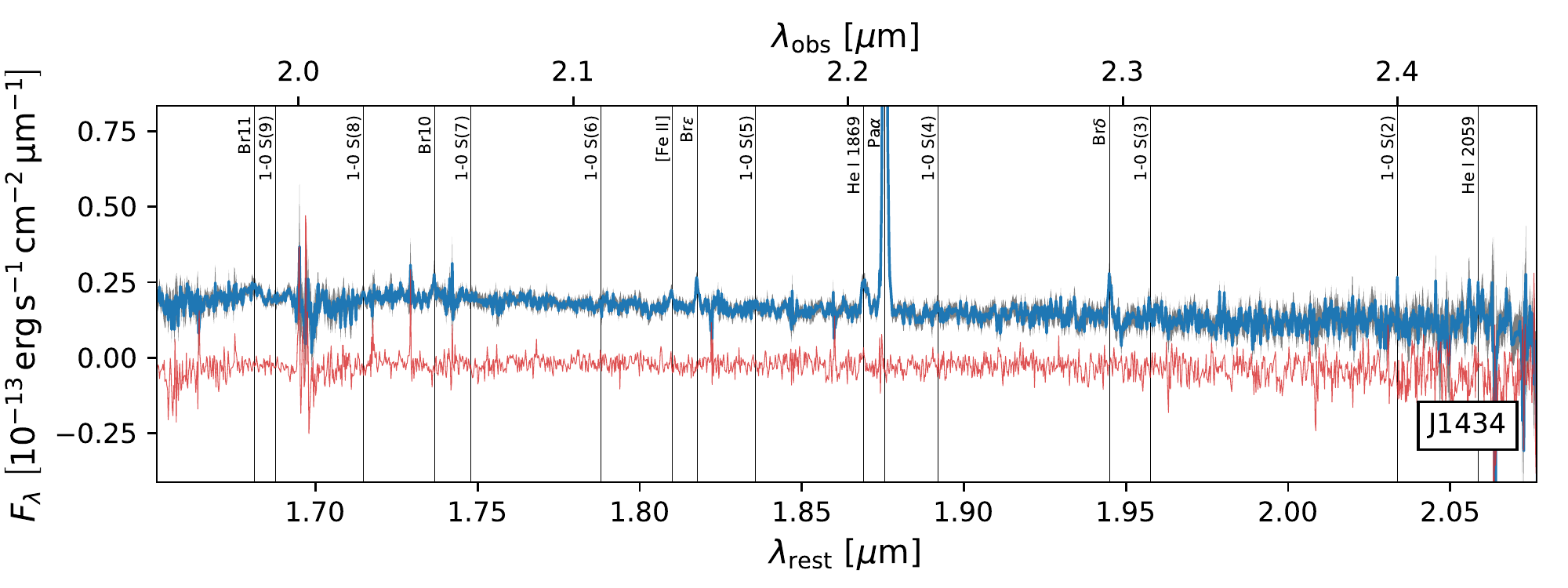}
	\caption{(Continued on next page)}
\end{figure*}
\begin{figure*}[ht!]
	\plotone{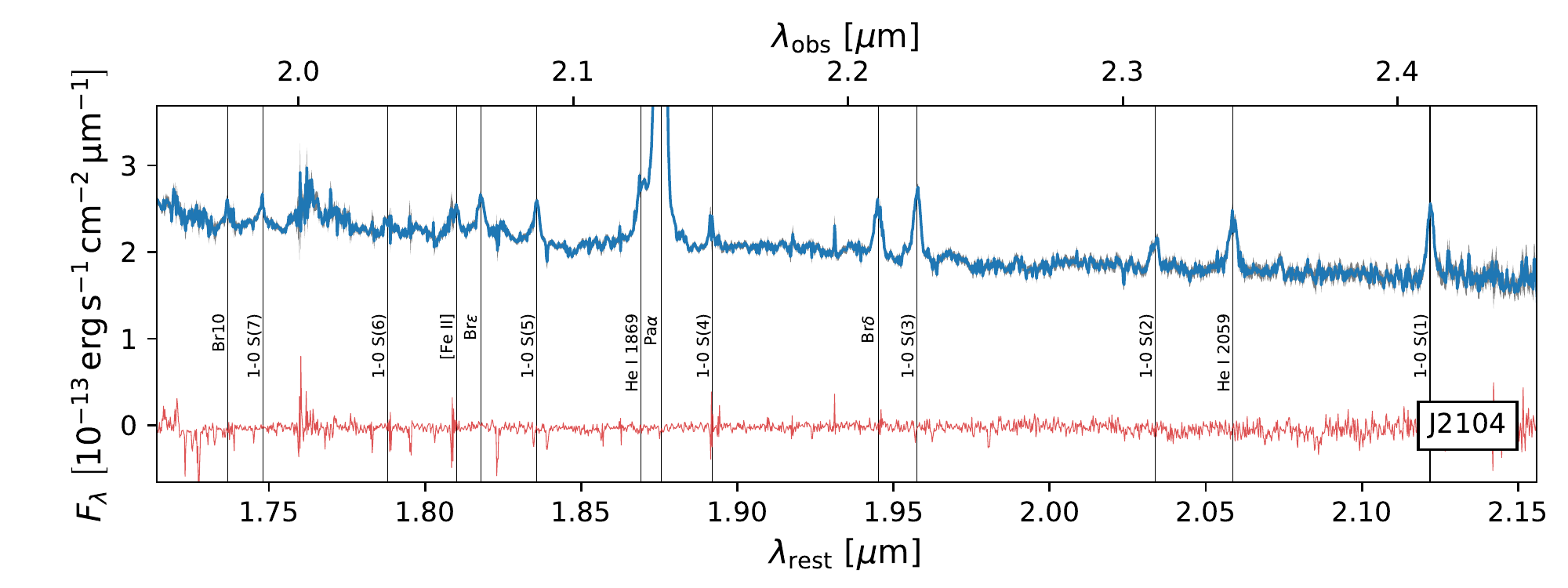}	
	\plotone{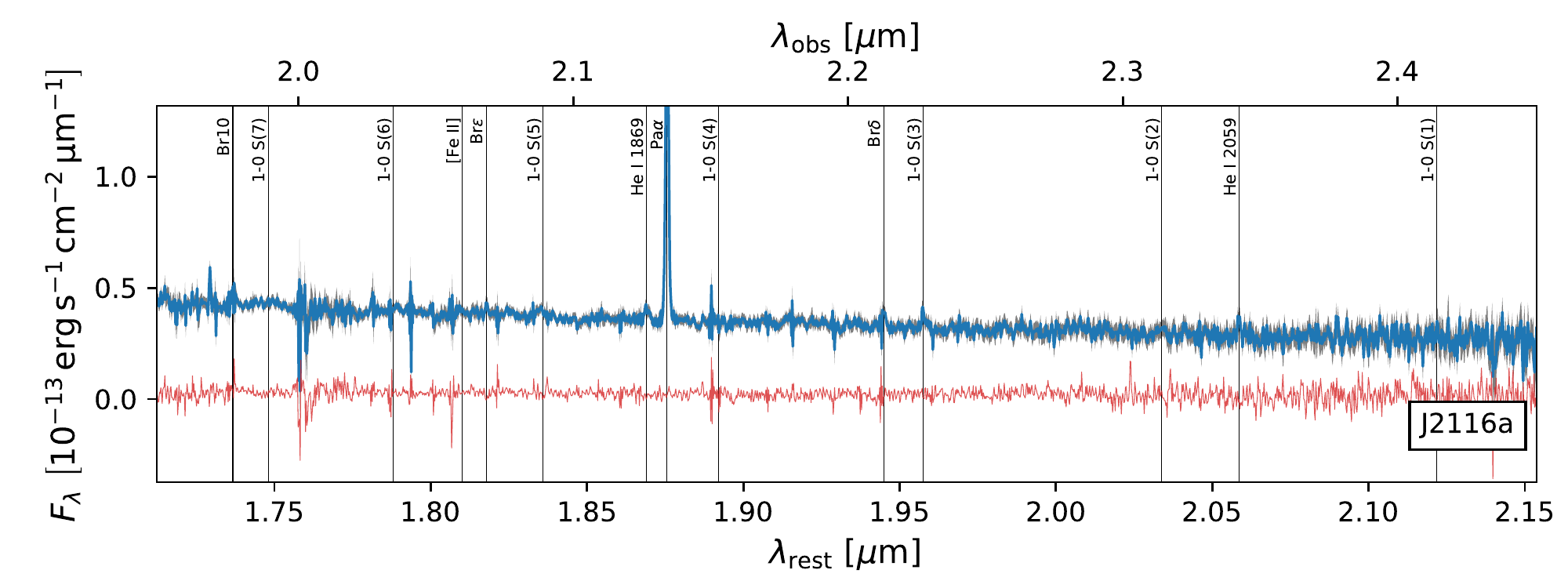}
	\plotone{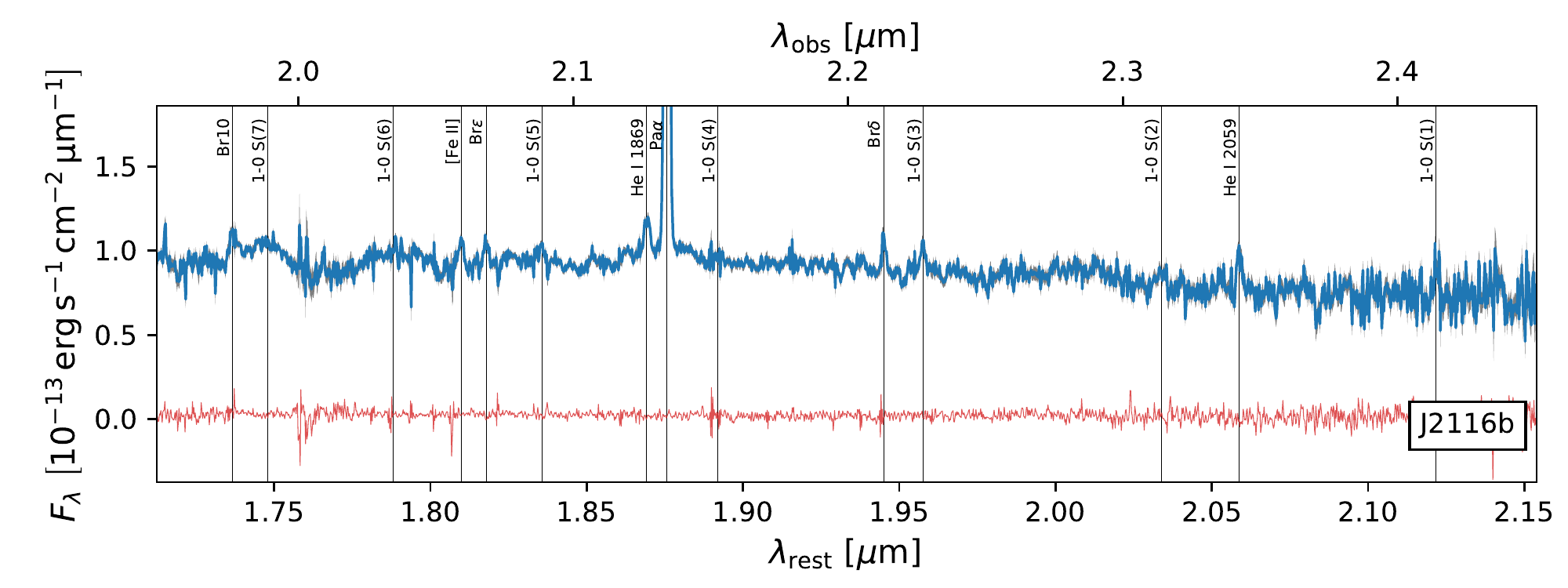}
	\addtocounter{figure}{-1}
	\caption{\label{fig:NIR-spectra}
		The NIR spectra for our sample extracted from two-pixel radius apertures centered on the peak pixel of Pa$\alpha$ integrated flux.
		The source spectrum is shown in blue, uncertainties are shown in gray, and the spectrum of the blank sky is shown in red.
	}
\end{figure*}

The SINFONI IFU data reveal a wealth of information about multiple gas phases of the ISM, which we can use to understand physical processes that drive galaxy evolution.
For each LBA, we extract a spectrum from a two-pixel radius aperture centered on the Pa$\alpha$ emission peak.
In these spectra, shown in Figure~\ref{fig:NIR-spectra}, we find bright recombination line features, primarily from neutral hydrogen (\ion{H}{1}) and neutral helium (\ion{He}{1}).
By measuring recombination line strengths and their ratios, we are able to infer properties of the ionized (\ion{H}{2}) regions, such as the nebular dust attenuation and the effective temperature of young stars.
We also detect molecular hydrogen ro-vibrational lines (H$_2$~1-0~S($\cdot$) series) in our spectra.
H$_2$ ro-vibrational line ratios are indicators of the gas excitation temperature and provide unique insight into feedback processes in the ISMs of LBAs.

\subsection{Dust attenuation}\label{sec:dust}

We model the effects of dust by assuming a uniform attenuating screen described by the \cite{1994ApJ...429..582C,2000ApJ...533..682C} attenuation curve, $k(\lambda)$.
This empirical curve relates the observed attenuation to the color excess, e.g., in the optical $V$-band: $A_V = k(V) \times E(B-V)$.
We characterize the color excess in nebular regions by using pairs of the brightest available hydrogen recombination lines (H$\beta$, H$\alpha$, and Pa$\alpha$).

We have obtained H$\alpha$ and H$\beta$ line fluxes by querying the SDSS DR14 SkyServer \citep{2018ApJS..235...42A}.
\cite{2004ApJ...613..898T} and \cite{2004MNRAS.351.1151B} describe the process by which these line fluxes are extracted from optical spectra over 3\arcsec{} SDSS fiber apertures.
For all LBAs except J2116, which has no H$\alpha$ flux available, we find $E(B-V)_{\rm neb}$ using the Balmer decrement \cite[assuming the attenuation curve from][]{2000ApJ...533..682C}.
The average nebular color excess for our sample is $\langle E(B-V)_{\rm H\alpha + H\beta} \rangle = 0.256 \pm 0.041$.

Our new SINFONI observations allow for additional estimates of the dust attenuation.
Using Pa$\alpha$ in the NIR data cubes along with H$\alpha$ and H$\beta$ lines enables robust estimation of the attenuation curve over a large range in wavelengths.
We extract Pa$\alpha$ over a 3\arcsec{} diameter aperture centered on the continuum emission, matching the SDSS spectroscopic fiber placement, and fit a single Gaussian component to each median-subtracted spectrum.
Integrated flux uncertainties are estimated by taking the variance of 10,000 MCMC samples, excluding a burn-in of 10,000 steps.
Using the \cite{2000ApJ...533..682C} attenuation curve and intrinsic line ratios from \cite{1987MNRAS.224..801H}, we compute unweighted averages for the LBA sample: $\langle E(B-V)_{\rm Pa\alpha + H\beta}\rangle = 0.316 \pm 0.113$ and $\langle E(B-V)_{\rm Pa\alpha + H\alpha}\rangle = 0.349 \pm 0.166$.
In Table~\ref{tab:attenuation}, we show $E(B-V)_{\rm neb}$ color excesses derived using these line ratios.

We find that there is some variation in nebular color excesses computed using different pairs of recombination lines, although the statistical significance is low due to large uncertainties.\footnote{
	We have reported unweighted averages for $E(B-V)_{\rm neb}$ thus far, because one source with unusually low uncertainties would otherwise dominate the average.
	Inclusion of J0212 in the weighted average would bias the sample $E(B-V)_{\rm neb}$ in an unrepresentative manner: $0.24 \pm 0.04$, $0.26 \pm 0.09$, and $0.24 \pm 0.11$ for H$\alpha$ + H$\beta$, Pa$\alpha$ + H$\beta$, and Pa$\alpha$ + H$\alpha$ respectively.
	When J0212 is excluded, the weighted average nebular excesses are $0.27 \pm 0.02$, $0.32 \pm 0.03$, and $0.35 \pm 0.04$ respectively.
}
The average nebular color excesses are related as $\langle E(B-V)_{\rm H\alpha+H\beta}\rangle < \langle E(B-V)_{\rm Pa\alpha+H\beta}\rangle < \langle E(B-V)_{\rm Pa\alpha+H\alpha}\rangle$, which is in line with expectations given that bluer photons are likely emitted from shallower regions in the case of a mixture  of dust, gas, and stars.
Pa$\alpha$ + H$\beta$-based attenuation spans the largest wavelength range and thereby offers the most sensitivity to differential attenuation.
Since Pa$\alpha$ and H$\beta$ lines are well-measured for all six LBAs, we use $E(B-V)_{\rm neb}$ estimated from Pa$\alpha$ + H$\beta$ to correct VLT/SINFONI nebular emission for dust attenuation.
This correction, using $\langle E(B-V)_{\rm Pa\alpha+H\beta} \rangle = 0.32 \pm 0.11$, is relatively small (but consistent with nebular color excesses from other LBG samples). 
For our sample, we find that correction for attenuation increases the NIR nebular line fluxes by $12-25\%$.
The dust corrections do not strongly impact line ratios; for example, the ratio of \ion{He}{1}~1869/Br$\delta$ is only increased by 2\%.

The continuum attenuation at 1530~\AA{}, $A_{1530}$, has been inferred from stellar population synthesis model fits to SDSS and \textit{GALEX} photometry (see \citealt{2005ApJ...619L..39S} for details).
Such a measurement depends most strongly on near- and far-UV observations, from which a UV slope can be derived (commonly parameterized by $\beta$; see, e.g., \citealt{1999ApJ...521...64M,2015ApJ...806..259R,2016ApJ...827...20S}).
For our sample of LBAs, \cite{2005ApJ...619L..35H} measure $A_{1530} \sim 2$, similar to $0 \lesssim A_{\rm FUV} \lesssim 2$ determined from Balmer lines (and assuming a \citealt{2001PASP..113.1449C} attenuation law; \citealt{2007ApJS..173..441H}), and comparable to $0 \lesssim A_{\rm FUV} \lesssim 3$ implied by the IRX-$\beta$ relation \citep[IRX$~\equiv L_{\rm IR}/L_{\rm UV}$;][]{2011ApJ...726L...7O}.
Individual uncertainties are not provided for model parameters, but the total scatter in $A_{1530}$ is about 30\% \citep[i.e., about 0.6 for our level of attenuation;][]{2005ApJ...619L..39S}.
Although some of this scatter is likely due to intrinsic scatter in a diverse sample of galaxies, we include an uncertainty of 0.5 in $A_{1530}$.
The far-UV attenuation can be converted to a continuum color excess, $E(B-V)_{\rm cont}$, assuming an attenuation curve similar to the one used for nebular regions \citep[with normalization $R_V = 4.05$;][]{2000ApJ...533..682C}.
In Table~\ref{tab:attenuation}, we show $E(B-V)_{\rm cont}$ derived from $A_{1530}$ under these assumptions.

We find that the ratio of UV continuum to nebular color excess is 
\begin{equation}
E(B-V)_{\rm cont} = (0.82 \pm 0.19) \times E(B-V)_{\rm H\alpha + H\beta},
\end{equation}
where in this case we show the nebular color excess derived from H$\alpha$ and H$\beta$ lines.
We can also compute
\begin{eqnarray}
E(B-V)_{\rm cont} =& (0.62 \pm 0.17) \times E(B-V)_{\rm Pa\alpha + H\beta},\\
E(B-V)_{\rm cont} =& (0.54 \pm 0.19) \times E(B-V)_{\rm Pa\alpha+ H\alpha},
\end{eqnarray}
using weighted averages of the color excess ratios.
In Table~\ref{tab:attenuation}, we show the continuum-to-nebular color excess ratio, hereafter defined as $f$, computed using Pa$\alpha$ and H$\beta$  for all six LBAs.

It is unsurprising that nebular light is more reddened than continuum light, as we have found here, because the ionized gas traces massive OB stars that tend to be embedded in high-attenuation birth clouds, while the continuum light originates from less massive stars that may have outlived or dissipated their birth clouds \cite[see, e.g.,][]{2000ApJ...539..718C}.
Simple models account for this effect by requiring two components of dust, one that is diffusely spread out across the entire ISM, and one that only heavily reddens or obscures the star-forming nebular regions.
If the two dust components are well-mixed throughout the galaxy, or at least in the regions traced by UV and emission lines, then the nebular and continuum color excesses should be equal to one another \citep[e.g.,][]{2001PASP..113.1449C}.
Further considerations such as complex geometries or morphological line-of-sight effects can be added to this model, but the overall intuition remains the same \citep[see, e.g.,][]{2011MNRAS.417.1760W,2013ApJ...775L..16K,2014ApJ...788...86P}.

For low-redshift, typical star-forming spirals, the ratio of continuum-to-nebular color excess is $f=0.44$ \citep{2000ApJ...533..682C}.
At higher redshifts, $f$ ranges from about one-half \citep[e.g.,][]{2013ApJ...779..135W,2014ApJ...788...86P} to near unity \citep[e.g.,][]{2013ApJ...777L...8K,2015ApJ...806..259R}.
The \cite{2000ApJ...533..682C}-like low values of $f$ can be physically interpreted using the previously mentioned two-component dust model, where the attenuated UV emission is not co-spatial with dustier nebular regions. 
Systems for which $f$ is higher may host a more homogeneous mixture of star-forming clouds and older stars, such that both UV-bright and ionizing stars are effectively attenuated by the same dust screen.

For our sample of LBAs, we find that $f_{\rm H\alpha+H\beta} > f_{\rm Pa\alpha+H\beta} > f_{\rm Pa\alpha+H\alpha}$ (although uncertainties are large).
This ordering of color excess ratios is expected because the color excess derived using longer-wavelength lines (i.e., Pa$\alpha$) is probing deeper into attenuated regions, whereas the shorter-wavelength Balmer lines can only probe shallower surfaces of star-forming clouds.
$A_V$ measured using Pa$\alpha$ is nearly an order of magnitude larger than for H$\beta$.
At the optical depths probed by H$\alpha$ emission, the ionizing stars are nearly as obscured as the non-ionizing stars that dominate the UV emission, based on the measured $f_{\rm H\alpha+H\beta} \sim 0.8$.
At the greater optical depths probed by Pa$\alpha$, the more embedded ionizing stars are more highly obscured than the non-ionizing populations, as evidenced from the more Calzetti-like $f_{\rm Pa\alpha + H\alpha}\sim 0.5$.

Other studies of LBAs \citep[see, e.g.,][]{2007ApJS..173..457B} have found that Balmer decrement-derived line attenuations are low (i.e., $A_{V,\rm neb} \sim 1$), but that continuum-based attenuation is comparable to that seen in typical low-$z$ star-forming galaxies.
Implied nebular excess ratios are therefore high, and comparable to our measurement of $f_{\rm H\alpha + H\beta} \sim 0.8$.
These previous findings are consistent with the interpretation that LBA stellar populations \textit{traced by Balmer lines} are still fully embedded in their birth clouds.
However, we show that UV-bright regions are not well-mixed with nebular regions traced at NIR wavelengths.

At higher redshifts, there is evidence that galaxies are forming stars at higher rates, and also undergo mergers more frequently. 
Both high-$z$ LBGs and our sample of LBAs exhibit high velocity dispersions, which may be a consequence of ongoing star formation.
The combination of recent merger and star-formation activity could then explain a high degree of mixing between dusty clouds hosting ionizing stars and the more diffuse dust associated with populations of less massive stars \citep[e.g.,][]{2015ApJ...806..259R,2015ApJ...807..141P,2016A&A...586A..83P}.
Intense winds and strong ionizing flux associated with compact starbursting regions can also lower the covering fraction of neutral gas in their dusty ISMs \citep[e.g.,][]{2014Sci...346..216B,2016ApJ...828..108R}

Not all high-redshift studies have found evidence for highly mixed populations of ionizing and non-ionizing stars.
\cite{2018arXiv180500016T} report $f_{\rm H\alpha+H\beta} \sim 0.75$ using SED fits to the continuum color excess \citep[as do][]{2016ApJ...826..159S}, although the authors note that the ratio falls to $\sim 1/3$ when the SMC attenuation curve is applied \citep[as described in][]{2003ApJ...594..279G,2016ApJ...828..107R} instead of the \cite{2000ApJ...533..682C} curve.
The SMC curve is steeper at shorter wavelengths, and appears to better match the canonical conversion factor of $f = 0.44$, particularly for lower-mass systems.
The higher-mass SINS sample \citep{2009ApJ...706.1364F} is characterized by a lower, Calzetti-like $f$, which the authors cite as evidence for an additional level of attenuation toward nebular regions.
\cite{2011MNRAS.417.1760W} are able to explain this effect using a strong correlation between $f$ and sSFR, which can be physically interpreted as an inverse relationship between the age of the starburst and the degree of mixing between younger and older stellar populations.

Our results are also consistent with an empirical ``double Calzetti'' law \cite[see, e.g.,][]{2000ApJ...539..718C,2013ApJ...779..135W,2014ApJ...788...86P,2017MNRAS.472.1372L}, which combines a steep \cite{1994ApJ...429..582C} attenuation curve for nebular regions at shorter wavelengths (e.g., for Balmer lines), and a shallower curve at longer wavelengths (e.g., for Pa$\alpha$).
At longer wavelengths, the nebular color excess appears higher because the optical depth reaches unity deeper into the dusty star-forming clouds where stars form.
The extra attenuation toward the most embedded clouds accounts for why ionized gas traced by Pa$\alpha$ is closer to the canonical \cite{2000ApJ...533..682C} ratio of $f \sim 0.44$.

\begin{deluxetable*}{l c ccc c}
	\tablewidth{0pt}
	\tablecolumns{6}
	\tabletypesize{\scriptsize}
	\tablecaption{Nebular and continuum color excesses \label{tab:attenuation}}
	\tablehead{
		\colhead{LBA} & 
		\colhead{$E(B-V)_{\rm cont}$} &
		\multicolumn{3}{c}{$E(B-V)_{\rm neb}$} &
		\colhead{Color excess ratio}
		\vspace{-0.75em}\\
		\colhead{} &
		\colhead{UV} &
		\colhead{H$\alpha$ + H$\beta$} & 
		\colhead{Pa$\alpha$ + H$\beta$} &
		\colhead{Pa$\alpha$ + H$\alpha$} &
		\colhead{$f_{\rm Pa\alpha + H\beta}$}
	}
	\startdata
    J0150  & $0.144$ & $0.251 \pm 0.003$ & $0.331 \pm 0.006$ & $0.366 \pm 0.006$ & $ 0.43 \pm  0.15$ \\
    J0212  & $0.183$ & $0.182 \pm 0.003$ & $0.132 \pm 0.006$ & $0.111 \pm 0.005$ & $ 1.39 \pm  0.38$ \\
    J0213  & $0.293$ & $0.271 \pm 0.012$ & $0.521 \pm 0.038$ & $0.630 \pm 0.040$ & $ 0.56 \pm  0.10$ \\
    J1434  & $0.243$ & $0.273 \pm 0.006$ & $0.299 \pm 0.011$ & $0.312 \pm 0.010$ & $ 0.81 \pm  0.17$ \\
    J2104  & $0.201$ & $0.303 \pm 0.004$ & $0.317 \pm 0.009$ & $0.325 \pm 0.009$ & $ 0.63 \pm  0.16$ \\
    J2116  & $0.203$ & $\cdots$ & $0.295 \pm 0.015$ & $\cdots$ & $ 0.69 \pm  0.17$ \\
    \hline
	Average & $0.211$ & $0.256 \pm 0.041$ & $0.316 \pm 0.113$ & $0.349\pm 0.166$ & $0.62 \pm 0.17$ \\
	\enddata
	\tablecomments{
		$E(B-V)_{\rm cont}$ is computed from UV continuum attenuation \citep{2005ApJ...619L..39S,2005ApJ...619L..35H} under the assumption of a \cite{2000ApJ...533..682C} attenuation curve with $R_V = 4.05$, and uniform uncertainty of $0.049$.
		$E(B-V)_{\rm neb}$ is computed from pairs of recombination lines, where H$\alpha$ and H$\beta$ line fluxes and uncertainties are obtained from the \cite{2004ApJ...613..898T} GalSpecLine catalog using the SDSS DR14 SkyServer, and Pa$\alpha$ fluxes are fit to SINFONI spectra extracted from $3\arcsec$ apertures to match SDSS fibers.
		We also show the ratio of continuum-to-nebular (Pa$\alpha$ + H$\beta$) color excess for each LBA.
		Averages and standard deviations for the LBA sample are also included, where the $E(B-V)_{\rm neb}$ and $f$ averages are unweighted and weighted by the inverse variance, respectively.
	}
\end{deluxetable*}

\subsection{The near-infrared recombination spectrum} \label{sec:recombination}
In addition to strong Pa$\alpha$ lines, we find hydrogen Brackett series lines and \ion{He}{1} $2.059~\micron{}$ and $1.869~\micron{}$ doublet lines present in all LBA spectra.
These lines arise from the H and \ion{He}{1} recombination cascades in gas photoionized by star formation or other physical processes.
In J2116, multiple lines have been detected unambiguously in both components, so we report results for the two separately.

\subsubsection{Measuring line fluxes and kinematics}
\begin{figure*}
	\plotone{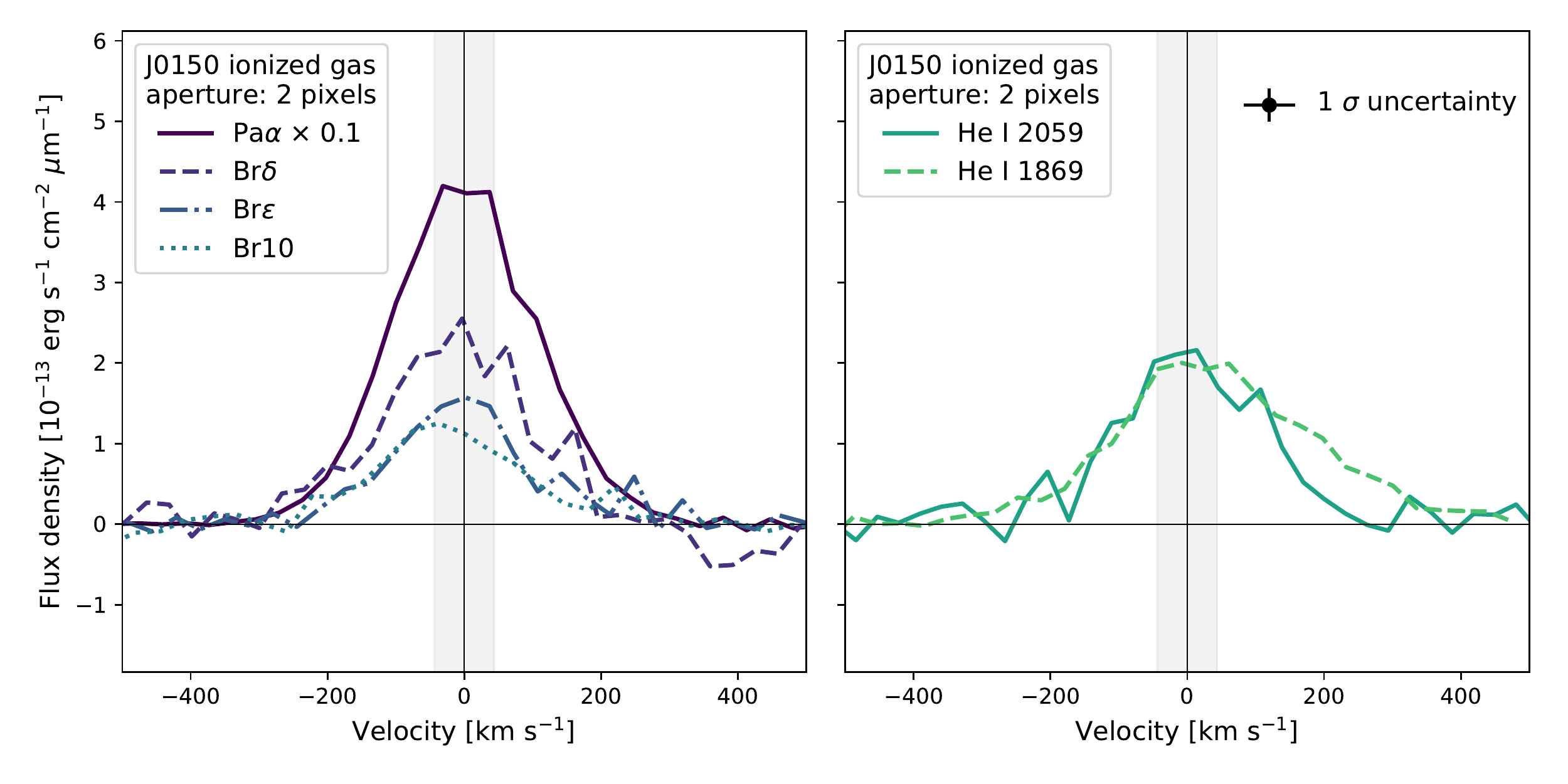}
	\caption{\label{fig:J0150-recombination-spectra}
		Recombination lines in J0150 for hydrogen (\textit{left}) and \ion{He}{1} (\textit{right}) relative to systemic velocity.
		Strong Pa$\alpha$ emission has been multiplied by $0.1$ for better viewing.
		Typical 1~$\sigma$ uncertainties are shown in the upper right corner of the right panel, and the FWHM spectral resolution is shown as a gray shaded region.
	} 
\end{figure*}

We measure the properties of recombination lines over a smaller aperture ($\sim 0.6\arcsec$, or about the size of the PSF) in order to accurately determine kinematics near the LBAs' central regions.
Spectra are extracted from a 2-pixel radius aperture centered on the brightest Pa$\alpha$ pixel as determined from the zeroth moment map.
For Pa$\alpha$, Br$\delta$, Br$\epsilon$, Br$10$, and \ion{He}{1} 2.059~$\micron{}$, we shift spectral lines to a common velocity frame and extract the spectra between $-1500~{\rm km~s^{-1}} < v < 1500~{\rm km~s^{-1}}$.
Br11 is observable for J0213 and J1434, but the line is also very faint, so we do not include it in our analysis.
For the case of the \ion{He}{1} doublet comprising lines at 1.869 and 1.870~\micron{}, due to strong Pa$\alpha$ emission at slightly longer wavelengths, we use the spectrum from $-2500~{\rm km~s^{-1}}< v < 500~{\rm km~s^{-1}}$.\footnote{Hereafter, we refer to \ion{He}{1} 2.059~$\micron{}$ as \ion{He}{1}~2059, and the \ion{He}{1} doublet at 1.869 and 1.870~\micron{} as \ion{He}{1}~1869, or \ion{He}{1}~1869+1870 when space permits.}
For each spectral ``cutout'' over its limited velocity range, we estimate the median absolution deviation (MAD) and convert to a standard deviation, $\sigma$, assuming Gaussian statistics.
We subtract the local continuum by taking a median of the (mostly) line-free channels $|v| > 500~{\rm km~s^{-1}}$ for each spectral cutout.
In Figure~\ref{fig:J0150-recombination-spectra}, we show recombination lines from an example median-subtracted, dust-uncorrected spectrum.

Each line is fit to a three-parameter Gaussian curve using SciPy optimization, and we sample parameter uncertainties using MCMC \citep[the \texttt{emcee} package;][]{emcee} with 10,000 steps (after 10,000 burn-in steps).
\ion{He}{1}~1869+1870 lines are blended, so we model them jointly with a double Gaussian at a single velocity and with fixed line ratios determined by \cite{1999ApJ...514..307B}, such that only three parameters are free to vary.
For Pa$\alpha$, we find that a single Gaussian does not always adequately fit the data, as residuals (data $-$ model) sometimes show signs of another Gaussian component.
We therefore also jointly fit two Gaussian components to Pa$\alpha$, and select between one-component and two-component models on the basis of a lower Akaike Information Criterion \citep[AIC;][]{1974ITAC...19..716A}.
We have implemented the following priors: (1) the line width\footnote{The FWHM line width is related to the standard deviation assuming a Gaussian distribution: FWHM~$= 2\sqrt{2\ln 2} \sigma$.} is assumed to be distributed $\sim \mathcal N({\rm FWHM}_{\rm Pa\alpha}, [50~{\rm km~s^{-1}}]^2)$ (i.e., centered on the Pa$\alpha$ FWHM line width measured from the clipped moment maps in Table~\ref{tab:kinematics}, and with standard deviation ${50~\rm km~s^{-1}}$), and (2) the velocity center is assumed to be distributed $\sim \mathcal N (v_{\rm sys}, [50~\rm km~s^{-1}]^2)$, where $v_{\rm sys}$ is the systemic velocity corresponding to the redshift.
For Pa$\alpha$, we allow parameters of the second component to freely vary, but with only a prior on the velocity center $\sim\mathcal N(v_{\rm sys}, [150~{\rm km~s^{-1}}]^2)$.

\begin{figure*}
	\plotone{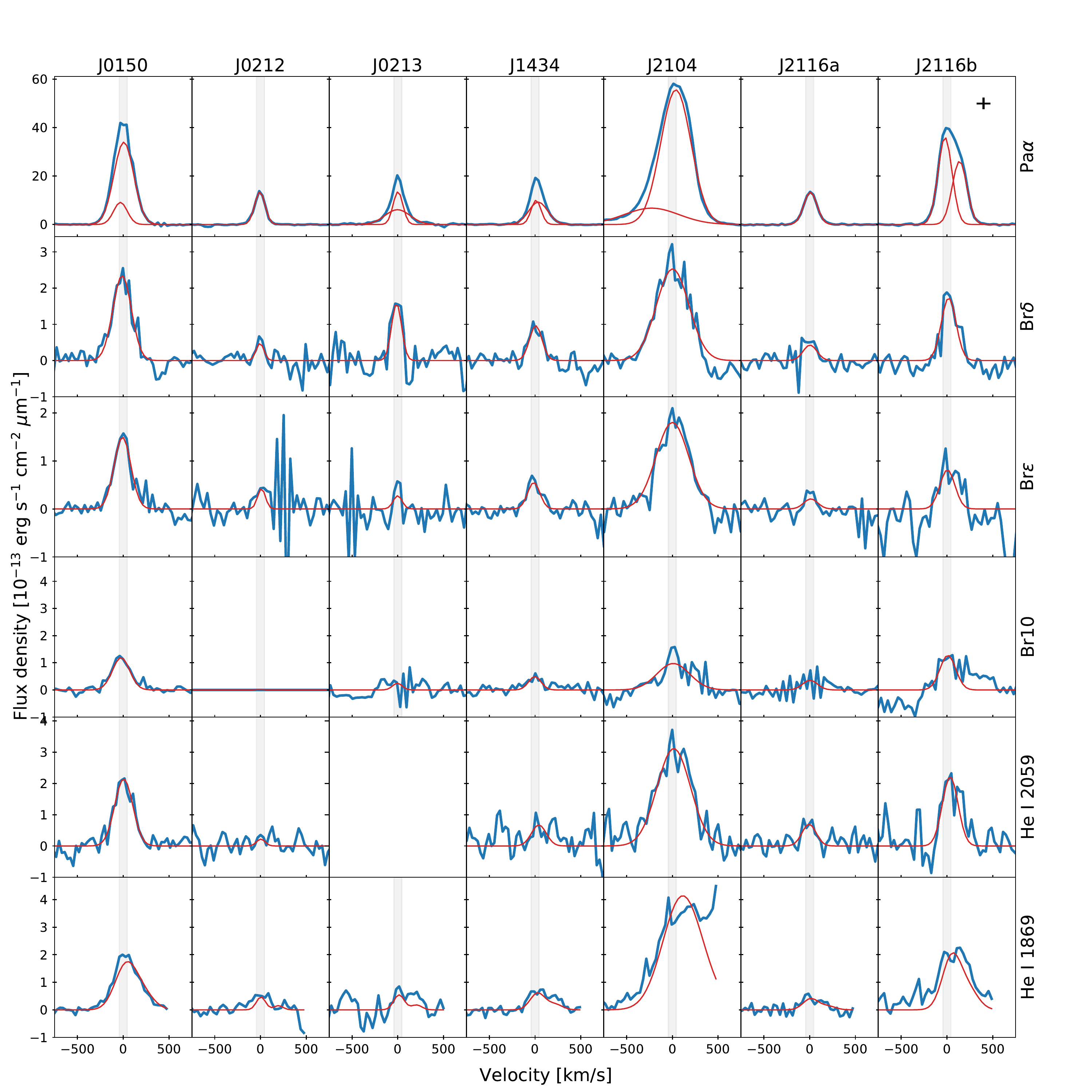}
	\caption{\label{fig:recombination-line-fits}
		Recombination lines (rows labeled at right) shown for each LBA (columns labeled at top).
		Spectra centered on the systemic velocity and extracted over a two-pixel radius aperture are shown in blue, and Gaussian model component fits are shown in red.
		Uncertainties representing $2\times$ the rms noise for Pa$\alpha$ and $2\times$ the spectral resolution are shown in the upper right corner, and gray shaded regions represent the FWHM spectral resolution.
	} 
\end{figure*}

In Figure~\ref{fig:recombination-line-fits}, we compare best-fit model recombination line components (shown in red) with the data (blue). 
For high-SNR lines, the Gaussian models fit the data well, and the model scatter is consistent with uncertainties in the data.
Some lines are fit with considerable uncertainty, such as the fainter lines from J0212 and J2116a.
For J0150, J2104, and J2116b, all available recombination lines are well-described by the model.

\subsubsection{Hydrogen recombination lines} \label{sec:recombination-line-results}

Best-fit line fluxes and widths are determined from the medians of posterior MCMC samples, and uncertainties from the standard deviations of these distributions.
In Table~\ref{tab:recombination-line-fluxes}, we show the total recombination line fluxes.
The values for Pa$\alpha$ are different from the fluxes in Table~\ref{tab:attenuation}, which were used for computing dust attenuation and meant to be compared directly with SDSS-measured H$\alpha$ and H$\beta$ lines \citep{2004ApJ...613..898T}. 
In this case, we have fit Pa$\alpha$ with either a single or double Gaussian line profile depending on which fit minimizes the AIC.
We note again that these spectra are extracted from a 2-pixel (or $\sim 0\farcs$6) aperture, whereas in Section~\ref{sec:dust}, the 3\arcsec{} SDSS matched-aperture spectra capture ionized gas dynamics over larger fractions of the galaxies and may include the outer disk or neighboring components.
The present apertures are also now centered on the Pa$\alpha$ integrated flux maxima, rather than the continuum brightness peaks.

In Table~\ref{tab:recombination-line-widths}, we show the recombination line widths.
We also list individual fluxes, line centers, and line widths for Pa$\alpha$ components in order to compare dynamics.
For the two LBAs with the faintest Pa$\alpha$ emission, J0212 and J2116a, only a single kinematic model component is favored, but the others require two components.
For most LBAs, the line width of the more luminous Pa$\alpha$ component is similar to those of the other recombination lines.
This more luminous component can be interpreted as tracing the star-forming ISM in each galaxy, with a velocity width that depends on the host's individual dynamics.
The fainter component can be attributed to ionized gas being heated and driven out of the galaxy disk, perhaps by supernovae, stellar winds, or AGN feedback, albeit with different degrees of confidence for different systems.
For J2104, the most Pa$\alpha$-luminous LBA, we find evidence for blueshifted ($v \approx -300~\rm km~s^{-1}$) ionized gas with significant Pa$\alpha$ flux, consistent with expectations for an energetic outflow.
For J0150, J0213, and J1434, the second Pa$\alpha$ component has a broader line width than the first.
While J2116b appears to have a bright blueshifted component, from inspecting Figure~\ref{fig:moment-maps}, we determine that a significant portion of the redshifted emission is actually off the detector, so the asymmetric line profile is not due to any physical phenomenon.

Other works have observed intense feedback and outflows in LBA systems.
For instance, \cite{2014Sci...346..216B} use \textit{HST}/COS to observe the FUV spectrum, including the Lyman continuum (LyC), of a radio-detected LBA.
The authors propose that radiation and shock heating from starbursts, evidenced by P-Cygni profiles in high-ionization transitions, can ionize gaps/holes in the neutral gas.
Supporting results from \cite{2015ApJ...810..104A} demonstrate that radiation feedback and starburst-driven winds are able to excavate ionized channels of gas.
This is particularly true for especially the most massive LBAs, which tend to host dominant compact objects \citep[DCOs; such as J0150, J0213, and J2104;][]{2011ApJ...730....5H}.
However, blueshifted Ly$\alpha$ emission---a proxy for escaping LyC emission---was \textit{not} significantly detected in J2104, despite our finding that it has the strongest evidence for extreme radiation and outflows in our sample.
Instead, the Ly$\alpha$ velocity profile resembles that of other (more typical) high-$z$ star-forming galaxies \citep[i.e., showing absorption and sometimes weak blueshifted emission;][]{2003ApJ...588...65S}.

\begin{figure}
	\plotone{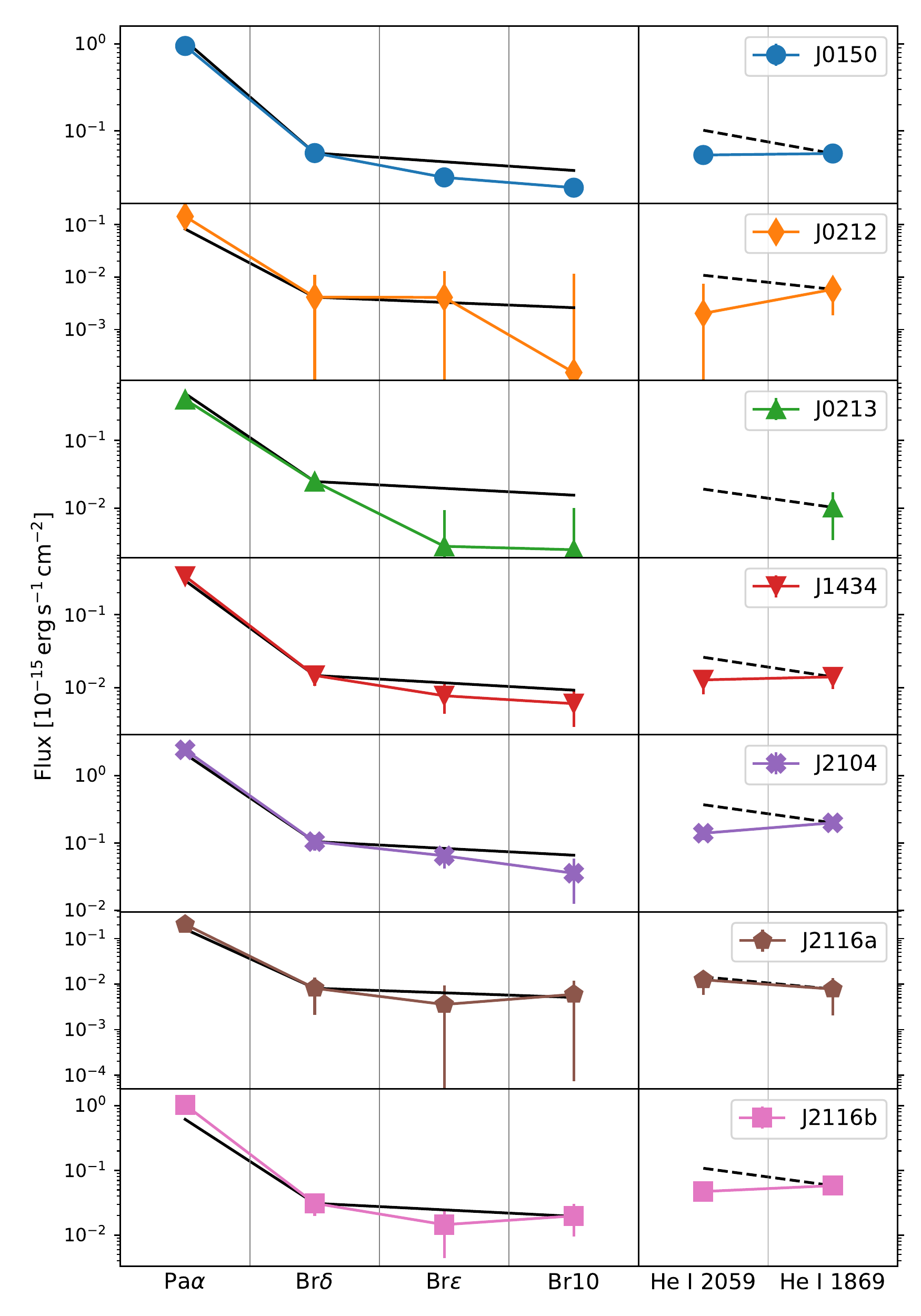}
	\caption{\label{fig:recombination-line-ratios}
		Dust-corrected hydrogen and \ion{He}{1} recombination line fluxes, normalized to Br$\delta$ and \ion{He}{1}~1869 respectively, shown for all LBAs in different colors and markers.
		Case B recombination line ratios are shown in black for hydrogen (solid; $T_e = 10^4$~K, $n_e = 10^3~\rm cm^{-3}$) and helium (dashed; $T_e = 10^4$~K, $n_e = 10^4~\rm cm^{-3}$).
		The \ion{He}{1} lines do not follow their Case B ratios, likely due to the combined effects of resonant scattering, dust or hydrogen absorption, and collisional population of the upper level for the 2.059~\micron{} line (see Section~\ref{sec:helium-recombination} for more).
		}
\end{figure}

In Figure~\ref{fig:recombination-line-ratios}, we compare observations to the dust-free emissivity ratios provided by \cite{1987MNRAS.224..801H} and \cite{1999ApJ...514..307B} for Case B recombination line ratios at $T_e = 10^4~\rm K$ for hydrogen ($n_e = 10^3~{\rm cm}^{-3}$) and neutral helium ($n_e = 10^4~{\rm cm}^{-3}$) respectively.
In the figure, fiducial hydrogen lines fluxes are normalized to the observed Br$\delta$ flux, and helium lines to \ion{He}{1}~1869+1870.
The recombination spectrum is generally not sensitive to the electron temperature or density of the ionized medium.
We find that case B predictions generally agree with the data.

Br$\epsilon$ and Br10 appear to be under-luminous for J0150, J0213, and J1434, whose data otherwise seem to match the Case B models.
Three possible explanations can conceivably contribute to this phenomenon.
First, the Br$\delta$ flux may be overestimated due to spectral features (of unknown origin) that artificially diminish the continuum (as can be seen in Figure~\ref{fig:recombination-line-fits}), and thereby elevate Br$\delta$ flux.
If this were indeed the case, then all lines would appear to be lowered with respect to Br$\delta$, \textit{except} Pa$\alpha$, which is fit using a two-component model and may include emission not captured by the other single-Gaussian fits.
Second, the nebular dust attenuation as estimated from a global integration over the large aperture might be considerably lower than attenuation extracted over only the central regions.
If this were the sole reason for the low Br$\epsilon$ and Br10 fluxes, then Pa$\alpha$, which is blueward of Br$\delta$, should also be slightly biased low; however, we can see that this is not the case in Figure~\ref{fig:recombination-line-ratios}.
Third, night sky lines (whose corrections are evident in Figure~\ref{fig:recombination-line-fits} as ringing patterns in the spectra) may add noise to the line fits, although we do not see any reason why the fluxes of only Br$\epsilon$ and Br10 would be biased low.

A combination of the first and second points can explain why Pa$\alpha$ and Br$\delta$ are apparently more luminous relative to Br$\epsilon$ and Br10 than expected for Case B recombination.
Our method of estimating the continuum generally produces valid results, but it is possible that unknown absorption lines exist near Br$\delta$.
Moreover, Pa$\alpha$ properties cannot be directly compared to those of other hydrogen recombination lines, due to the former's different fitting procedure.
Finally, radial gradients in metallicity may correlate with decreased attenuation in the outer regions of the LBAs \citep[see, e.g.,][]{2010Natur.467..811C}.
This trend can lead to underestimates of $E(B-V)$ in central regions, and thus to underprediction of the unattenuated fluxes of bluer lines (e.g., Br10 and Br$\epsilon$).

\begin{deluxetable*}{l rrrrrrr}
	\tablewidth{0pt}
	\tablecolumns{8}
	\tablecaption{Recombination line fluxes \label{tab:recombination-line-fluxes}}
	\tablehead{
		\colhead{LBA} &
		\colhead{Pa$\alpha$\tablenotemark{a}} &
		\colhead{Br$\delta$} &
		\colhead{Br$\epsilon$} &
		\colhead{Br$10$} &
		\colhead{\ion{He}{1}~$2059$} &
		\colhead{\ion{He}{1}~$1869$\tablenotemark{b}}
	}
	\startdata
    J0150  & $9.53 \pm 0.98$ & $0.56 \pm 0.05$ & $0.28 \pm 0.05$ & $0.22 \pm 0.04$ & $0.52 \pm 0.05$ & $0.54 \pm 0.05$ \\
    J0212  & $1.43 \pm 0.07$ & $0.04 \pm 0.06$ & $0.04 \pm 0.08$ & $0.00 \pm 0.23$ & $0.03 \pm 0.05$ & $0.06 \pm 0.04$ \\
    J0213  & $4.01 \pm 0.64$ & $0.25 \pm 0.07$ & $0.03 \pm 0.09$ & $0.02 \pm 0.08$ & \nodata & $0.13 \pm 0.07$ \\
    J1434  & $3.34 \pm 0.57$ & $0.15 \pm 0.04$ & $0.08 \pm 0.04$ & $0.06 \pm 0.03$ & $0.14 \pm 0.05$ & $0.14 \pm 0.05$ \\
    J2104  & $24.07 \pm 1.57$ & $1.03 \pm 0.28$ & $0.66 \pm 0.25$ & $0.40 \pm 0.22$ & $1.42 \pm 0.29$ & $1.96 \pm 0.29$ \\
    J2116a & $2.05 \pm 0.11$ & $0.07 \pm 0.06$ & $0.03 \pm 0.06$ & $0.05 \pm 0.05$ & $0.13 \pm 0.07$ & $0.06 \pm 0.06$ \\
    J2116b & $10.29 \pm 1.21$ & $0.30 \pm 0.11$ & $0.14 \pm 0.10$ & $0.21 \pm 0.11$ & $0.43 \pm 0.13$ & $0.56 \pm 0.14$ \\
	\enddata
	\tablecomments{
		Dust-corrected flux ($10^{-15}~\rm erg~s^{-1}~cm^{-2}$) fit using Gaussian models for each recombination line.            
		Uncertainties have been determined through sampling using MCMC.
		For cases in which the central wavelengths are shifted out of the LBA spectrum, no data are shown ($\cdots$).    
	}
	\tablenotetext{a}{Pa$\alpha$ line fit using one or two summed Gaussian components.}
	\tablenotetext{b}{\ion{He}{1} 1869+1870 blended lines fit using two Gaussians with fixed flux ratio, relative velocity, and equal line width.}
\end{deluxetable*}

\begin{deluxetable*}{l rrr c rrr rrrrrr}
	\tablewidth{0pt}
	\tablecolumns{14}
	\tablecaption{Recombination line widths \label{tab:recombination-line-widths}}
	\tablehead{
		\colhead{LBA} &
		\multicolumn{3}{c}{Pa$\alpha$ (1)} &
		\colhead{} &
		\multicolumn{3}{c}{Pa$\alpha$ (2)} &
		\colhead{Br$\delta$} &
		\colhead{Br$\epsilon$} &
		\colhead{Br$10$} &
		\colhead{\ion{He}{1}~$2059$} &
		\colhead{\ion{He}{1}~$1869$} \\
		\cline{2-4} \cline{6-8}        
		\colhead{} &
		\colhead{Flux} &
		\colhead{Center} &
		\colhead{Width} &
		\colhead{} &
		\colhead{Flux} &
		\colhead{Center} &
		\colhead{Width} &
		\colhead{} &
		\colhead{} &
		\colhead{} &
		\colhead{} &
		\colhead{}
	}
	\startdata
    J0150  & $ 8.0 \pm 0.7$ & $  6$ & $ 255 \pm    6$ & & $ 1.6 \pm 0.7$ & $-30$ & $ 178 \pm   15$ & $ 237 \pm   22$ & $ 212 \pm   30$ & $ 221 \pm   33$ & $ 229 \pm   24$ & $ 278 \pm   27$ \\
    J0212  & $ 1.4 \pm 0.1$ & $ -8$ & $ 131 \pm    5$ & & \nodata & \nodata & \nodata & $  98 \pm   52$ & $ 104 \pm   57$ & $  72 \pm   59$ & $ 111 \pm   52$ & $ 120 \pm   50$ \\
    J0213  & $ 1.8 \pm 0.5$ & $  2$ & $ 123 \pm   16$ & & $ 2.2 \pm 0.4$ & $ -3$ & $ 312 \pm   46$ & $ 123 \pm   28$ & $ 119 \pm   52$ & $ 129 \pm   52$ & \nodata & $ 139 \pm   45$ \\
    J1434  & $ 2.0 \pm 0.4$ & $ 34$ & $ 249 \pm   21$ & & $ 1.3 \pm 0.4$ & $ 12$ & $ 130 \pm   13$ & $ 166 \pm   34$ & $ 154 \pm   41$ & $ 161 \pm   45$ & $ 185 \pm   48$ & $ 188 \pm   39$ \\
    J2104  & $19.8 \pm 1.2$ & $ 39$ & $ 391 \pm   13$ & & $ 4.3 \pm 1.0$ & $-235$ & $ 700 \pm  172$ & $ 414 \pm   45$ & $ 421 \pm   50$ & $ 423 \pm   49$ & $ 431 \pm   42$ & $ 483 \pm   42$ \\
    J2116a & $ 2.0 \pm 0.1$ & $  8$ & $ 169 \pm    6$ & & \nodata & \nodata & \nodata & $ 180 \pm   47$ & $ 171 \pm   48$ & $ 190 \pm   49$ & $ 192 \pm   47$ & $ 184 \pm   45$ \\
    J2116b & $ 5.9 \pm 0.8$ & $-21$ & $ 179 \pm   12$ & & $ 4.4 \pm 0.8$ & $138$ & $ 188 \pm   17$ & $ 196 \pm   42$ & $ 186 \pm   45$ & $ 206 \pm   43$ & $ 196 \pm   38$ & $ 254 \pm   41$ \\
	\enddata
	\tablecomments{
		FWHM line widths in $\rm km~s^{-1}$ for each recombination line. 
		For Pa$\alpha$, the dust-corrected flux ($10^{-15}~\rm erg~s^{-1}~cm^{-2}$), central velocity ($\rm km~s^{-1}$ from the systemic velocity), and
		FWHM ($\rm km~s^{-1}$) are shown for each component.
		For lines that are shifted out of the LBA spectra, or if no second Pa$\alpha$ component is fit, we report no data ($\cdots$).    
	}
\end{deluxetable*}

\subsubsection{\ion{He}{1} recombination} \label{sec:helium-recombination}

We can constrain the spectral hardness of ionizing radiation by comparing line ratios across the \ion{He}{1} and hydrogen recombination cascades.
Such an approach has been shown to be a useful diagnostic of the effective temperature ($T_{\rm eff}$) of ionizing star populations \citep[e.g.,][]{1984ApJ...284..118G,1992ApJ...397..117D,1995MNRAS.277..577D,2001ApJ...552..544F,2004ApJ...606..237R}.
We predict LBA \ion{He}{1} and \ion{H}{1} line ratios by using Cloudy photoionization models \citep[version 17.00;][]{2017RMxAA..53..385F}, for which $T_{\rm eff}$ can be an input parameter.

\begin{figure*}
	\plotone{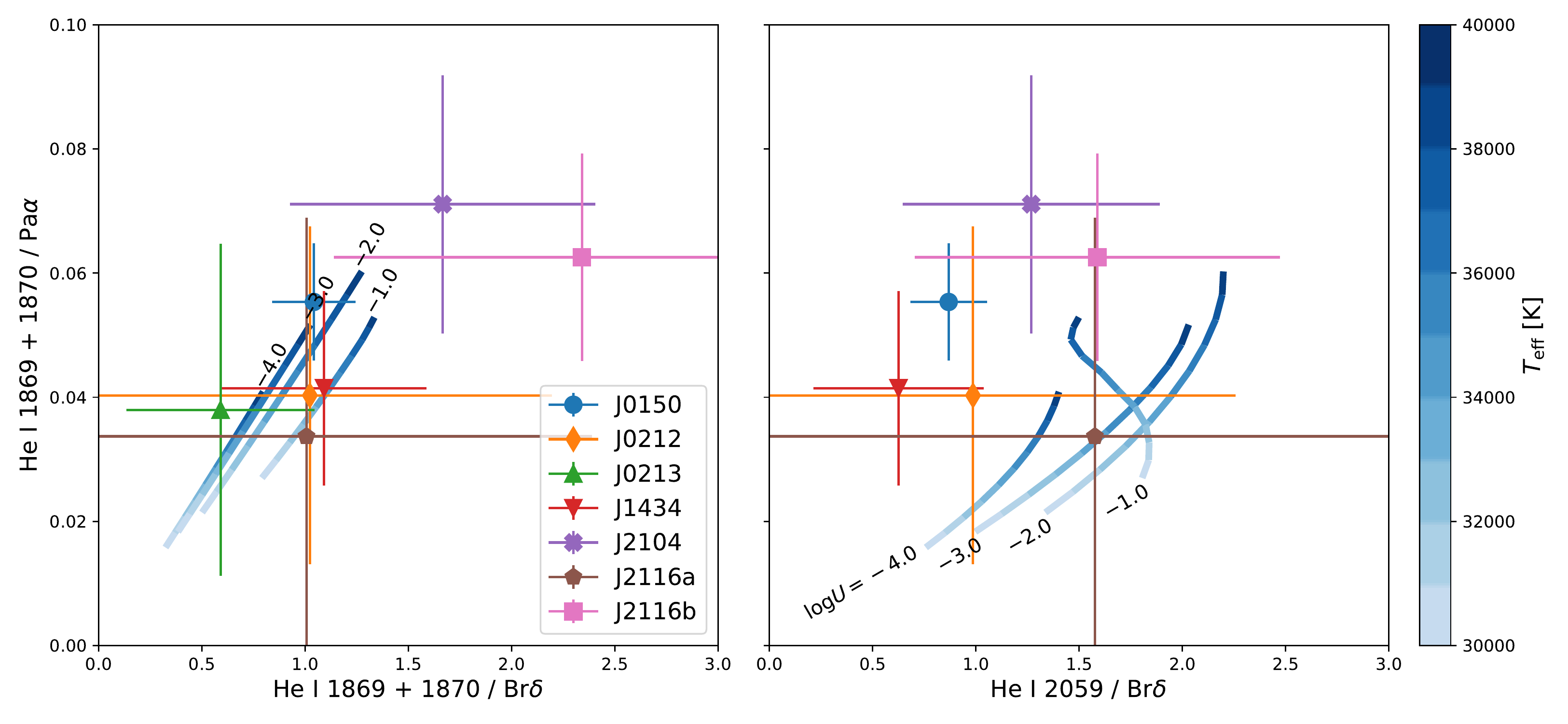}
	\caption{\label{fig:helium-hydrogen-ratios}
		Observed dust-corrected \ion{He}{1}-to-\ion{H}{1} recombination line ratios (points) shown with Cloudy model predictions (curves).
		We show model results while varying effective temperature (colorbar) and logarithmic ionization parameter $\log U \in \{-4.0, -3.0, -2.0, -1.0\}$, as labeled.
	}
\end{figure*}

Our model assumes a spherical geometry of only hydrogen and helium (with solar He/H ratio) surrounding a central star with effective temperature in the range $30,000~{\rm K}\leq T_{\rm eff} \leq 40,000~{\rm K}$.
We also vary the logarithmic ionization parameter, $-4.0 \leq \log U \leq 0$ (where $U$ is the ratio of ionizing photon density to hydrogen density), and assume Case B recombination.
For our LBA sample, we find [\ion{O}{3}]/[\ion{O}{2}] flux ratios close to unity \citep[][]{2004ApJ...613..898T}, which implies that $\log U \sim -3$ \citep[see, e.g.,][]{2015AA...576A..83S}.
In Figure~\ref{fig:helium-hydrogen-ratios}, we present measurements and model predictions for \ion{He}{1}/\ion{H}{1} recombination line ratios.

We find that the \ion{He}{1}~2059 recombination line flux is lower than expected from Case B predictions (see, e.g., Figure~\ref{fig:recombination-line-ratios}).
In an ionized, dust-free medium, resonant pumping of the $1\,^1S \rightarrow 2\,^1P$ transition via an optically thick 584~\AA{} line amplifies the production of \ion{He}{1}~2059 photons.
However, dust attenuation and neutral hydrogen absorption in realistic nebulae reduce fluorescent \ion{He}{1}~2059 emission \citep[e.g.,][]{2003MNRAS.340..799L}.
Collisional depopulation from $2\,^3S \rightarrow 2\,^1P$ and ionization fractions can also impact the production of \ion{He}{1}~2059 photons (see \citealt{1989agna.book.....O,1993ApJ...419..181S} for more detailed discussions).
Therefore, we do not use the \ion{He}{1}~2059 line to constrain the effective blackbody temperature of O and B stars.
For completeness, however, we still plot \ion{He}{1}~1869+1870/Pa$\alpha$ vs. \ion{He}{1}~2059/Br$\delta$ in the right panel of Figure~\ref{fig:helium-hydrogen-ratios}.

\ion{He}{1}~1869+1870/Pa$\alpha$ and \ion{He}{1}~1869+1870/Br$\delta$ flux ratios are shown in the left panel of Figure~\ref{fig:helium-hydrogen-ratios}.
The combination of line ratios, which both increase monotonically with $T_{\rm eff}$, can be used to estimate the effective stellar temperatures.
Most LBAs have $T_{\rm eff} \lesssim 40,000{~\rm K}$, consistent with those of typical low-$z$ star-forming galaxies.
For J2104 and J2116b, the LBAs with highest Pa$\alpha$ fluxes, we also infer extremely high effective temperatures (i.e., $T_{\rm eff} > 40,000$~K for $\log U = -3$).\footnote{We repeat our comparison for \ion{He}{1} and \ion{H}{1} lines extracted over larger apertures (5~pixel radius, or $\sim 1.5\arcsec$), and find that J2104 has anomalously high \ion{He}{1}/\ion{H}{1} line ratios as before (implying $T_{\rm eff} > 40,000$~K), whereas J2116b becomes more consistent with $T_{\rm eff} = 40,000$~K.}

Recent works have reported that the effective temperature of newly formed stars in higher-$z$ systems may be much higher than in the nearby Universe, with $T_{\rm eff} \sim 50,000-60,000$~K \citep[see, e.g.,][]{2014ApJ...795..165S}.
A top-heavy IMF, in which the proportion of newly formed massive and high-$T_{\rm eff}$ stars is higher than traditional models \citep[such as][]{1955ApJ...121..161S,2001MNRAS.322..231K,2003PASP..115..763C}, may account for the increased $T_{\rm eff}$ seen in J2104 and J2116b.
A top-heavy IMF in LBGs is also consistent with theoretical work supporting the formation of very massive ($> 100~M_\sun$) stars in the early Universe when few or no metals are present \citep[see, e.g.,][]{2013RPPh...76k2901B,2018MNRAS.479.4544M}, although  our LBAs have gas-phase metallicities similar to those of other $z \sim 2-3$ LBGs that exhibit enhanced nebular excitation (e.g., \citealt{2017ApJ...836..164S}).

A related possibility is that J2104 hosts a significant fraction of massive star binaries, similar to stellar populations that are theorized to reside in high-$z$ star-forming galaxies but are noticeably absent in typical low-$z$ systems (see, e.g., \citealt{2016ApJ...826..159S}).
Such massive star binaries could also help explain some of the Wolf-Rayet features found in J2104's optical spectrum (\citealt{2008A&A...485..657B}) as well as in other LBAs (e.g., \citealt{2014Sci...346..216B}).
The stellar evolution of massive binaries may include an extended ($\sim 100$~Myr) phase during which stellar $T_{\rm eff}$ can reach as high as $\sim 100,000$~K \citep[to the point of doubly ionizing helium;][]{2012MNRAS.419..479E,2017PASA...34...58E}.
High-mass X-ray binaries are known to be more prevalent in distant low-metallicity systems \citep[i.e., from deep stacks of $2-10$~keV $z \leq 4$ LBGs;][]{2013ApJ...762...45B}.
Nearby low-metallicity LBAs are also characterized by elevated X-ray luminosity per SFR \citep[][]{2013ApJ...774..152B}, reinforcing the possibility that high effective temperatures in our LBA sample are due to massive star binaries.

Some LBAs appear to be starburst/AGN composite systems based on optical spectral diagnostics \citep{2009ApJ...706..203O}, and their ionization properties are elevated but not extreme for the low-redshift Universe \citep[e.g.,][]{2009MNRAS.399.1191C,2016ApJ...822...62B}. 
The properties of DCOs in the centers of some LBAs are consistent with modest nuclear activity, which may also elevate $T_{\rm eff}$.
In particular, there is strong evidence that J2104 hosts low-luminosity (obscured) AGN \citep[based on X-ray detections;][]{2011ApJ...731...55J}; however, its nucleus remains undetected in very long baseline radio observations \citep[][]{2012MNRAS.423.1325A}.
It is possible to attribute the hard ionizing spectrum ($T_{\rm eff} > 40,000$~K) and X-ray emission to an AGN, although we would then expect supermassive black hole accretion to affect \ion{He}{1}/\ion{H}{1} for other members of our sample (i.e., J1434 and J0150), which is not the case.
Indeed, AGN in LBAs are relatively weak and do not dominate their energy output \citep{2011ApJ...731...55J}.
In Section~\ref{sec:line-ratios}, we discuss warm molecular and ionized gas line ratios, which will provide further evidence that feedback from massive stars is the most important source for ISM heating in LBA systems.

\subsubsection{[\ion{Fe}{2}] recombination}
We at least marginally detect [\ion{Fe}{2}]~1.810~\micron{} line emission for all LBAs except J0212.
[\ion{Fe}{2}] traces low-ionization gas, since the electronic potential is only 7.9~eV.
For most nearby galaxies, [\ion{Fe}{2}] 1.257 or 1.644~\micron{} emission can be used to characterize shocks driven by supernovae \citep{1989A&A...214..307O,2003AJ....125.1210A} or AGN processes in Seyfert galaxies \citep[see, e.g.,][]{2009MNRAS.394.1148S}.
We detect [\ion{Fe}{2}]~1.810~\micron{} at $>3~\sigma$ significance for only J0150, and at $\sim 1.5~\sigma$ for J0213, J2104, and J2116b.
Due to its faintness, we do not include it in our tables and figures.

\subsection{H$_2$ ro-vibrational spectrum} \label{sec:ro-vibrational}

Molecular hydrogen can emit ro-vibrational lines in the NIR when heated or fluorescently excited.
In local star-forming galaxies, these H$_2$ emission lines are often attributed to shocks driven by young stars, or to photon-dominated regions (PDRs) near UV radiation (or in extreme cases, in X~ray-dominated regions; XDRs).
Gas can be heated by collisional excitation in both PDRs and shocks \citep[e.g.,][]{1989ApJ...342..306H,1991ApJ...383..205W}, while PDRs can also be responsible for fluorescent excitation of H$_2$ \citep[e.g.,][]{1987ApJ...322..412B,1989ApJ...338..197S}.
The properties of warm H$_2$ emission are important for understanding star formation feedback and its effects on the ISM, which can impact the host galaxy's evolution.
However, H$_2$ ro-vibrational emission lines are redshifted into the mid-infrared at high redshifts, and thus remain poorly studied for traditional samples of LBGs.
Our SINFONI observations provide us with the unique opportunity to measure multiple H$_2$ ro-vibrational lines for LBAs, and the potential to transfer this knowledge to their higher-redshift cousins.

We use the same single Gaussian fitting procedure as before to measure ro-vibrational line fluxes and widths, and run 10,000-step MCMC simulations to estimate uncertainties.
Extracted H$_2$ line fluxes and line widths are shown in Tables~\ref{tab:rovibrational-line-fluxes} and \ref{tab:rovibrational-line-widths}.
For the NIR recombination lines, we correct for extinction using the \cite{2000ApJ...533..682C} nebular gas prescription as described in Section~\ref{sec:dust}, as appropriate for star-forming clouds.

The column density of molecular hydrogen in each upper ro-vibrational state is proportional to the observed line flux, and from the distribution of lines a gas excitation temperature, $T_{\rm ex}$, can be inferred.
Following \cite{1978ApJ...223..464B}, we compute the column density of molecular hydrogen by measuring surface brightness $I = (1/\Omega) \int S_\lambda d\lambda$ using a circular aperture with two-pixel radius, with
\begin{equation}
N_{\rm H_2} = \frac{4\pi I}{A_{ul} h\nu},
\end{equation}
where $A_{ul}$ is the Einstein coefficient \citep[provided by][]{1977ApJS...35..281T} and $h\nu$ is the photon energy.
We divide the column density by the appropriate rotational level statistical weight and compare with upper ro-vibrational level energies \citep[as measured by][]{1984CaJPh..62.1639D}.

\begin{figure}
	\plotone{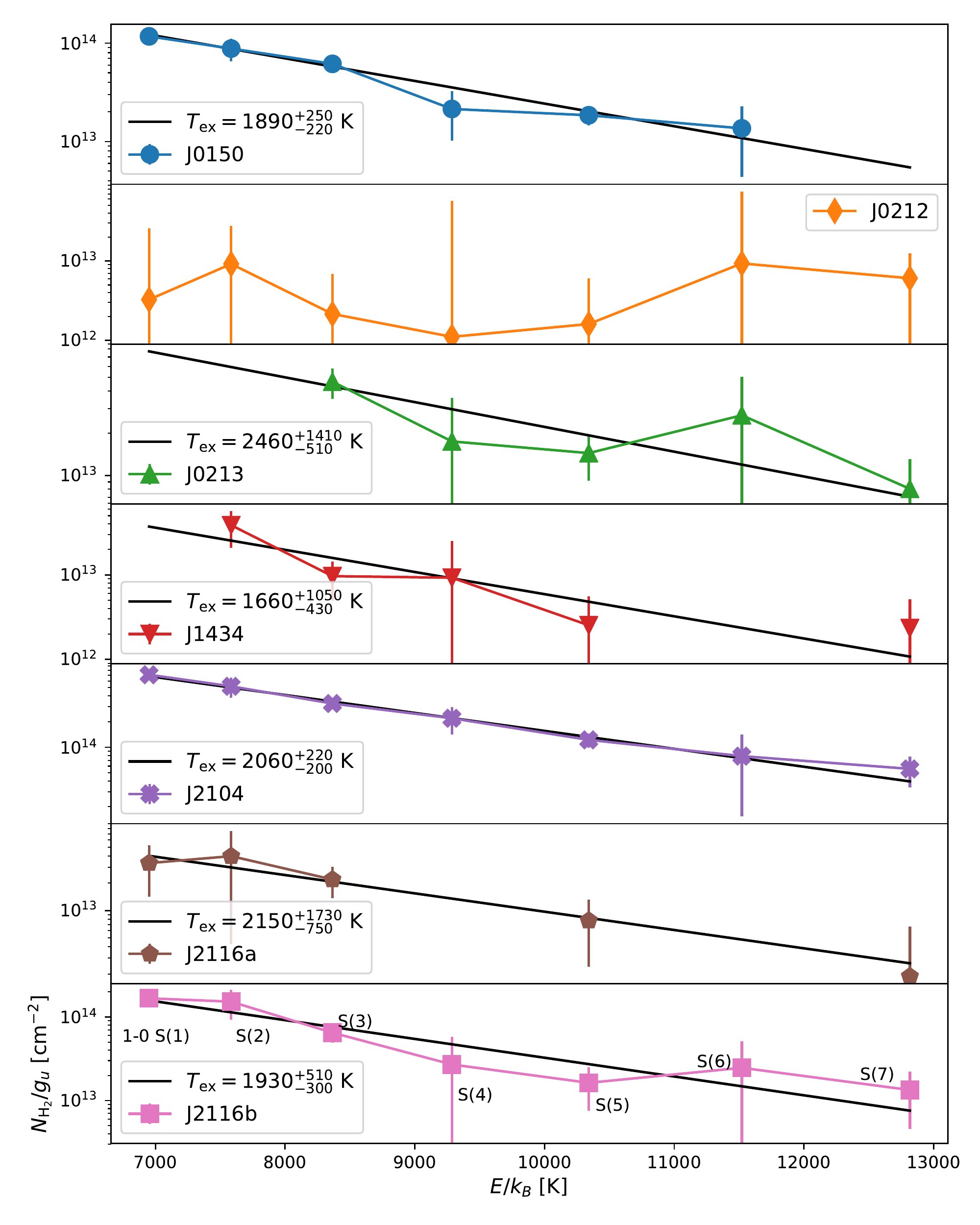}
	\caption{\label{fig:rovibrational-line-ratios}
		Dust-corrected column density normalized by level degeneracy for
		H$_2$ $v = 1$ upper rotational levels plotted against energy in 
		temperature units (color). 
		A single best-fit excitation temperature ($T_{\rm ex}$) is also plotted (in black) for each LBA component aside from J0212, whose ro-vibrational spectrum lacks high enough SNR to fit a temperature profile.
	}
\end{figure}

We then fit a single-temperature model to the logarithmic column density and estimate uncertainties using a MCMC routine as before.\footnote{Because we have used a logarithmic scale for the column density of molecular gas, $N_{\rm mol}$, we approximate uncertainties as $\sigma_{\ln N_{\rm mol}} \approx \sigma_{N_{\rm mol}}/N_{\rm mol}$. This linear approximation breaks down when $\sigma_{N_{\rm mol}} \gtrsim N_{\rm mol}$ and may cause bad fits for low-SNR lines, but we find that our fits are qualitatively consistent with the data.}
Our fit assumes that the H$_2$ gas is optically thin and adequately described by a single temperature.
H$_2$ gas becomes optically thick at column density $\sim 10^{26}~{\rm cm}^{-2}$ \citep{1978ApJ...223..464B}; our LBAs are well below this threshold ($N_{\rm H_2} \lesssim 10^{15}~{\rm cm}^{-2}$).
Other studies have found that multiple temperatures are sometimes needed to fit the ro-vibrational spectrum \citep[see, e.g.,][]{2000A&A...354..439K}, especially when higher energy transitions reveal the need for a hotter component.
For the range of energies probed by our SINFONI observations, single-temperature fits describe the H$_2$ data well.
Figure~\ref{fig:rovibrational-line-ratios} shows the ro-vibrational lines and best-fit models assuming a 3:1 ortho-to-para H$_2$ ratio \cite[see also][]{1999ApJ...516..371S}.

The H$_2$ 2-1~S($\cdot$) transitions can be useful for distinguishing between shocks and UV fluorescence \citep[e.g.,][]{1987ApJ...322..412B}.
High ratios of 2-1~S(1)/1-0~S(1) $\gg 0.1$ are inconsistent with shock heating \citep{1982ARA&A..20..163S}, while lower values may originate from shocked gas \textit{or} dense PDRs \citep[e.g.,][]{1990ApJ...365..620B,1997ARA&A..35..179H}.
However, we do not detect any H$_2$~2-1 S($\cdot$) emission lines in our spectra.
H$_2$~2-1~S(1) is always redshifted out of our spectral range, but the 2-1~S(3) line (and/or other bright $v=2$ lines) should be detectable in at least some of our LBAs even if the $v=2-1$ transitions are not probing even warmer H$_2$. 
Thus, our non-detection of H$_2$~2-1~S$(\cdot)$ lines is consistent with both UV heating and shocks as origins for the warm H$_2$ emission.

Bright [\ion{Fe}{2}] emission has been invoked as evidence for shocks \citep[e.g.,][]{1988A&A...203..278M,1989ApJ...342..306H,2001A&A...369L...5O}, but it requires extremely widespread shocks with high pre-shock densities.
\cite{1996ApJ...466..561M} find that observed [\ion{Fe}{2}] emission may not require shocks after all, but may instead be straightforwardly produced in XDRs \citep[e.g., due to irradiation from massive stars or AGN;][]{2005A&A...436..397M}.
XDR models have been able to successfully reproduce both the [\ion{Fe}{2}] and H$_2$ ro-vibrational emission observed in nearby AGN.
For our LBA sample, [\ion{Fe}{2}]~1.810~\micron{} emission lines are detected at $> 1.5~\sigma$ in 4/7 spectra, which lends credence to the idea that our LBA sample's H$_2$ lines represent X-ray-heated gas emission.
In Section~\ref{sec:line-ratios}, we will see that both [\ion{Fe}{2}] and H$_2$ emission are faint relative to hydrogen recombination lines, supporting a scenario in which recently formed massive stars (including high-mass X-ray binaries, but not AGN) are responsible for exciting the molecular gas.

Using a sample of LBAs similar to ours (with four overlapping members), \cite{Contursi+17} use \textit{Herschel}/PACS spectroscopy to measure [\ion{O}{1}]~63~\micron{} and [\ion{C}{2}]~158~\micron{} lines in order to probe physical conditions.
\cite{Contursi+17} favor PDRs over shocks for all detected LBAs in their sample on the basis of low [\ion{O}{1}]~63~\micron{}/[\ion{C}{2}]~158~\micron{} ratios \citep{1989ApJ...342..306H}.
\cite{2016MNRAS.457...64I} also come to the conclusion that, for a sample of low-metallicity blue compact dwarf galaxies, the dominant mechanism for exciting H$_2$ is fluorescence from strong UV radiation fields.

For our sample of LBAs, we find that the H$_2$ ro-vibrational emission is consistent with gas heating and line pumping from UV photons (and X-rays, in some cases).
Our observations are not inconsistent with molecular shocks, but we also do not find strong evidence to warrant the invocation of shocks.
Star formation-driven shocks are unlikely given the prodigious energy requirements for galaxy-wide H$_2$ emission.
Although tidal shocks may be present for interacting or merging systems \citep[e.g.,][]{2010MNRAS.406..535P,2015ApJ...807..134G}, LBAs' global warm molecular gas properties are not significantly different for merging and non-interacting systems.
Therefore, we interpret these findings as evidence that warm H$_2$ gas is primarily excited in (dense) PDRs for LBAs. 
Therefore, we favor the \cite{Contursi+17} interpretation of PDRs near star-forming regions as the primary locus of H$_2$ excitation.

\begin{deluxetable*}{l rrrrrrr}
	\tablewidth{0pt}
	\tablecolumns{8}
	\tablecaption{Ro-vibrational line fluxes \label{tab:rovibrational-line-fluxes}}
	\tablehead{\vspace{-0.5em}\\
		&
		\multicolumn7c{H$_2~$1-0 spectrum} \\
		\cline{2-8}        
		\colhead{LBA} &
		\colhead{S(1)} &
		\colhead{S(2)} &
		\colhead{S(3)} &
		\colhead{S(4)} &
		\colhead{S(5)} &
		\colhead{S(6)} &
		\colhead{S(7)}
	}
	\startdata
    J0150  & $  29.3 \pm   5.3$ & $  11.3 \pm   4.4$ & $  31.9 \pm   4.9$ & $   4.5 \pm   3.6$ & $  13.1 \pm   4.2$ & $   3.3 \pm   3.4$ & $ -35.5 \pm   6.6$ \\
    J0212  & $   0.8 \pm   8.5$ & $   1.2 \pm   3.6$ & $   1.1 \pm   3.7$ & $   0.2 \pm  17.7$ & $   1.1 \pm   4.8$ & $   2.3 \pm  24.4$ & $   4.3 \pm   7.0$ \\
    J0213  & \nodata & \nodata & $  24.2 \pm   9.0$ & $   3.7 \pm   5.8$ & $  10.2 \pm   5.7$ & $   6.6 \pm   8.8$ & $   5.7 \pm   5.4$ \\
    J1434  & \nodata & $   5.0 \pm   3.5$ & $   5.0 \pm   3.7$ & $   1.9 \pm   5.0$ & $   1.8 \pm   3.3$ & $  -1.6 \pm   5.8$ & $   1.7 \pm   3.0$ \\
    J2104  & $ 177.7 \pm  27.9$ & $  66.7 \pm  27.0$ & $ 168.6 \pm  29.2$ & $  45.9 \pm  24.7$ & $  86.4 \pm  24.9$ & $  19.3 \pm  23.6$ & $  39.5 \pm  23.8$ \\
    J2116a & $   8.3 \pm   7.2$ & $   5.1 \pm   6.9$ & $  11.4 \pm   6.5$ & $  -1.8 \pm   5.9$ & $   5.5 \pm   5.8$ & $  -0.4 \pm   5.7$ & $   1.3 \pm   5.1$ \\
    J2116b & $  41.8 \pm  13.7$ & $  19.5 \pm  11.6$ & $  33.5 \pm  12.3$ & $   5.7 \pm   9.7$ & $  11.5 \pm   9.4$ & $   6.0 \pm   9.8$ & $   9.5 \pm   9.5$ \\
	\enddata
	\tablecomments{
		Dust-corrected flux ($10^{-17}~\rm erg~s^{-1}~cm^{-2}$) fit using a single Gaussian model for each H$_2~1-0$ ro-vibrational line.            
		Uncertainties have been determined through sampling using MCMC.
		No data are shown ($\cdots$) if a line is shifted out of the LBA spectrum.    
	}
\end{deluxetable*}

\begin{deluxetable*}{l rrrrrrr}
	\tablewidth{0pt}
	\tablecolumns{8}
	\tablecaption{Ro-vibrational line widths \label{tab:rovibrational-line-widths}}
	\tablehead{\vspace{-0.5em} \\
		&
		\multicolumn7c{H$_2~$1-0 spectrum} \\
		\cline{2-8}        
		\colhead{LBA} &
		\colhead{S(1)} &
		\colhead{S(2)} &
		\colhead{S(3)} &
		\colhead{S(4)} &
		\colhead{S(5)} &
		\colhead{S(6)} &
		\colhead{S(7)}
	}
	\startdata
    J0150  & $ 235 \pm   31$ & $ 189 \pm   41$ & $ 243 \pm   33$ & $ 184 \pm   49$ & $ 218 \pm   42$ & $ 183 \pm   48$ & ? \\
    J0212  & $  98 \pm   54$ & $  99 \pm   49$ & $  95 \pm   51$ & $  94 \pm   60$ & $  88 \pm   55$ & $  83 \pm   59$ & $ 120 \pm   56$ \\
    J0213  & \nodata & \nodata & $ 188 \pm   50$ & $ 131 \pm   52$ & $ 124 \pm   42$ & $ 131 \pm   53$ & $ 143 \pm   47$ \\
    J1434  & \nodata & $ 136 \pm   56$ & $ 154 \pm   49$ & $ 146 \pm   53$ & $ 147 \pm   48$ & ? & $ 143 \pm   47$ \\
    J2104  & $ 407 \pm   45$ & $ 428 \pm   45$ & $ 422 \pm   45$ & $ 426 \pm   49$ & $ 423 \pm   46$ & $ 423 \pm   50$ & $ 426 \pm   49$ \\
    J2116a & $ 181 \pm   48$ & $ 186 \pm   51$ & $ 188 \pm   47$ & ? & $ 188 \pm   48$ & ? & $ 181 \pm   51$ \\
    J2116b & $ 204 \pm   43$ & $ 193 \pm   49$ & $ 200 \pm   44$ & $ 190 \pm   49$ & $ 189 \pm   47$ & $ 184 \pm   51$ & $ 186 \pm   50$ \\
	\enddata
	\tablecomments{
		H$_2~1-0$ ro-vibrational line widths ($\rm km~s^{-1}$).
		No data are shown ($\cdots$) if a line is shifted out of the LBA spectrum, and a question mark (?) is shown if the best-fit flux is negative.    
	}
\end{deluxetable*}

\section{Warm and cool molecular gas} \label{sec:H2 masses}

Comparisons of H$_2$ ro-vibrational lines with tracers of other phases of the ISM can help improve our understanding of how multiphase gas is heated in PDRs, winds, and/or outflows \citep[e.g.,][]{2009ApJ...700L.149V,2010MNRAS.405..898O,2013ApJ...775L..15R,2014Natur.511..440T,2016MNRAS.455.1830T,2017ApJ...837..149G,2017ApJ...843...18V,2018MNRAS.480L.111G}.
Large variations in the mass ratios of cool molecular gas, such as the $\sim 30$~K gas most easily traced by CO rotational observations, and warm gas traced by H$_2$ ro-vibrational emission, may support scenarios in which cool gas is dissociated and re-forms over longer timescales.
Correlations between the cool/warm ratio and driving mechanism (e.g., modest star-formation vs. vigorous starbursts vs. AGN) are also informative for discriminating between models.

\cite{1982ApJ...253..136S} consider thermally populated warm emitting H$_2$, and estimate its mass assuming optically thin emission. 
\citet[][and citations within their Section~4.8]{2009MNRAS.394.1148S} provide the formula for determining the warm H$_2$ mass at an assumed $T_{\rm ex}=2000~$K using H$_2$~1-0 S(1) flux:
\begin{equation} \label{eq:warm-H2}
M_{\rm H_2} = 5.078 \times 10^{13} ~M_\sun \left (\frac{F_{\rm H_2\,{\text 1-0}\,S(1)}}{\rm erg\,s^{-1}\,cm^{-2}}\right ) \left (\frac{D_L}{\rm Mpc}\right )^2.
\end{equation}
From Figure~\ref{fig:rovibrational-line-ratios} we see that $T_{\rm ex}=2000~$K is within 1~$\sigma$ of the best fits for most LBAs (although the H$_2$ excitation temperature is unconstrained for J0212).

In Table~\ref{tab:H2-masses} we compare warm and cool H$_2$ gas masses, where cool H$_2$ masses are calculated using low-resolution radio observations of the CO(1-0) line \citep{Contursi+17}, assuming $X_{\rm CO} = 2\times 10^{20} ~\rm  cm^{-2}~(K~km~s^{-1})^{-1}$ following \cite{2013ARAA..51..207B}.
\cite{Contursi+17} detect significant CO~(1-0) flux using the IRAM PdBI for four LBAs in our sample.
For J0150, we have used the CO~(1-0) flux obtained from CARMA observations with longer integration times \citep{2014MNRAS.442.1429G}.
We determine warm H$_2$ masses using Equation~\ref{eq:warm-H2}.
For LBAs where H$_2~$1-0 S(1) is redshifted out of the $K$-band, we compute the flux $F_{\rm H_2~{\text 1-0}~S(1)} \approx 1.9 \times F_{\rm H_2~{\text 1-0}~S(3)}$, where H$_2~$1-0 S(3) is the second-brightest observed H$_2$ line and the line ratio has been derived from a $T_{\rm ex}=2000~$K ro-vibrational spectrum.

J0213 appears to host the largest mass fraction of warm molecular gas, even higher than the vigorously star-forming J2104 (although the two agree to within 1~$\sigma$ uncertainties).
Could this indicate that a larger fraction of the molecular ISM is excited, due to either shocks or PDRs/XDRs?
Both J0213 and J2104 are composite AGN/star-forming galaxies according to their emission line ratios \citep{2009ApJ...706..203O} using the \citet[][or BPT]{1981PASP...93....5B} classification, so this seems possible.
As discussed above, the LBAs likely host low-mass obscured AGNs, based on follow-up X-ray and interferometric radio observations  \citep[][]{2011ApJ...731...55J,2012MNRAS.423.1325A,2013ApJ...774..152B}.

\begin{deluxetable}{l r r r}[ht!]
	\tablewidth{0pt}
	\tablecolumns{4}
	\tabletypesize{\scriptsize}
	\tablecaption{Warm and cool H$_2$ masses		\label{tab:H2-masses}}
	\tablehead{
		\colhead{Object} & 
		\colhead{$M_{\rm H_2}^{\rm warm}$ [$10^3~M_\sun$]} &
		\colhead{$M_{\rm H_2}^{\rm cool}$ [$10^{10}~M_\sun$]} &
		\colhead{$M_{\rm H_2}^{\rm warm}/M_{\rm H_2}^{\rm cool}$ [$10^{-7}$]}
		}
	\startdata
	J0150 & $7.2 \pm 1.3$ & $1.2 \pm 0.1$\tablenotemark{b} & $6.2 \pm 1.7$ \\
	J0212 & $\cdots$ & $\cdots$ & $\cdots$ \\
	J0213 & $27.5 \pm 10.2$\tablenotemark{a} & $1.9 \pm 0.2$\tablenotemark{c} & $15.1 \pm 5.9$\\
	J1434 & $3.7 \pm 2.7$\tablenotemark{a} & $1.3 \pm 0.4$\tablenotemark{c} & $2.8 \pm 2.3$ \\
	J1434 (total) & $9.4 \pm 3.0$\tablenotemark{a} & $1.3 \pm 0.4$\tablenotemark{c} & $7.2 \pm 3.2$ \\
	J2104 & $37.4 \pm 5.9$\tablenotemark{a} & $4.1\pm 0.3$ & $9.3 \pm 1.6$\\
	J2116a & $1.8 \pm 1.6$ & \nodata & \nodata \\
	J2116b & $8.9 \pm 2.9$ & \nodata & \nodata \\
	\hline
	IC 5063 & $0.82 \pm 0.12 $\tablenotemark{d} & $0.050 \pm  0.005$\tablenotemark{e} &$16 \pm 3$\\
	M82 & $12$\tablenotemark{f} & $0.13$\tablenotemark{g} & $92$ \\
	F11119+3257 & 52\tablenotemark{h} & 0.15\tablenotemark{i} & $340$ \\
	\enddata
	\tablecomments{Total warm and cool H$_2$ masses and mass ratios shown for our sample of LBAs and three nearby objects. 
		For J1434, we show H$_2$ derived from only its main component, and also from the sum of its main and companion as described in Section~\ref{sec:merger-shocks}.
		(a) In some cases, H$_2$~1-0~S(1) is unavailable, so we approximate the S(1) line flux with the S(3) flux using the method described in the text.
		Measurements are taken from
		{(b) }{\cite{2014MNRAS.442.1429G},}
		{(c) }{\citet{Contursi+17},}	
		{(d) }{\cite{2014Natur.511..440T},}
		{(e) }{\cite{2013AA...552L...4M},}
		{(f) }{\cite{2009ApJ...700L.149V},}
		{(g) }{\cite{2002ApJ...580L..21W},}
		{(h) }{\cite{2013ApJ...775L..15R},}
		and {(i) }{\cite{2014AA...562A..21C}.}
		}

\end{deluxetable}

We compare our results with those for three prototypical low-$z$ objects that have been well-studied in the literature: M82, a starburst galaxy; F11119+3257, a ULIRG hosting a Type 1 quasar; and IC 5063, a radio galaxy hosting a Type 2 Seyfert.
Their warm and cool molecular masses are shown in Table~\ref{tab:H2-masses}.
For each system, we report the total CO-derived cool H$_2$ mass; although it is possible to separate a broad CO component from a rotating disk for these nearby, bright, and well-resolved systems, we show the global masses in order to facilitate comparisons with our LBA sample.
Cool H$_2$ masses have been estimated using a Galactic CO-to-H$_2$ conversion factor, $\alpha_{\rm CO}$, for IC~5063 \citep{2013AA...552L...4M}.
A ULIRG-like $\alpha_{\rm CO}$ factor is appropriately adopted for F11119+3257 \citep{2014AA...562A..21C}, and a combination of starburst and Galactic $\alpha_{\rm CO}$ factors are used for different gas components of M82 \citep{2002ApJ...580L..21W}.

We find that IC 5063 best matches the LBA sample in terms of warm/cool H$_2$ mass ratio.
It is worth noting that the ionized gas (Br$\gamma$), warm molecular gas (H$_2$~1-0~S(1)), and cool molecular gas (CO(2-1)) components of IC~5063 are each kinematically distinct \citep{2014Natur.511..440T}, not unlike the multiphase gas components in our own sample.
IC~5063 is also notable among our three comparison objects in having a morphologically early type, in which one might expect star formation and thus warm gas to be centrally concentrated; moreover, it hosts a Type 2 AGN (which may be found in some of our LBAs). 
Whether feedback in LBAs is due to compact starbursts or to obscured AGN, the energetic photons are expected to originate from near the galaxy's nucleus.
For M82, which is undergoing a galaxy-wide starburst driving a large-scale wind, and F11119+3257, in which a merger-induced starburst is likely contributing to strong ${\rm H_2}$ emission across the system, the {\it global} warm-to-cool molecular gas mass ratios may be elevated in a way that is only the case for the center of IC~5063.

\section{Resolved near-infrared line ratios} \label{sec:line-ratios}

\begin{figure*}
	\includegraphics[width=\textwidth]{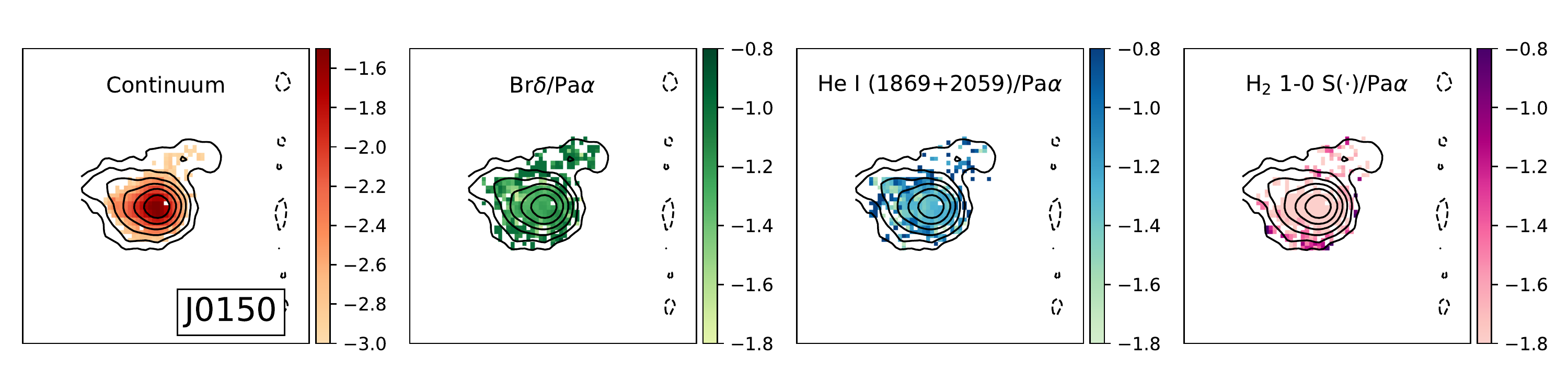}
	\includegraphics[width=\textwidth]{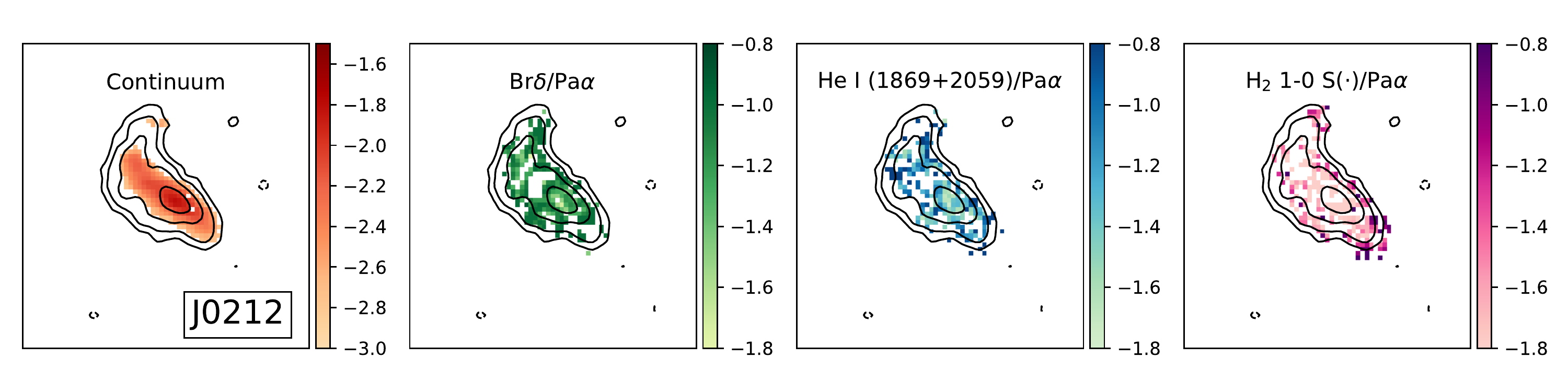}
	\includegraphics[width=\textwidth]{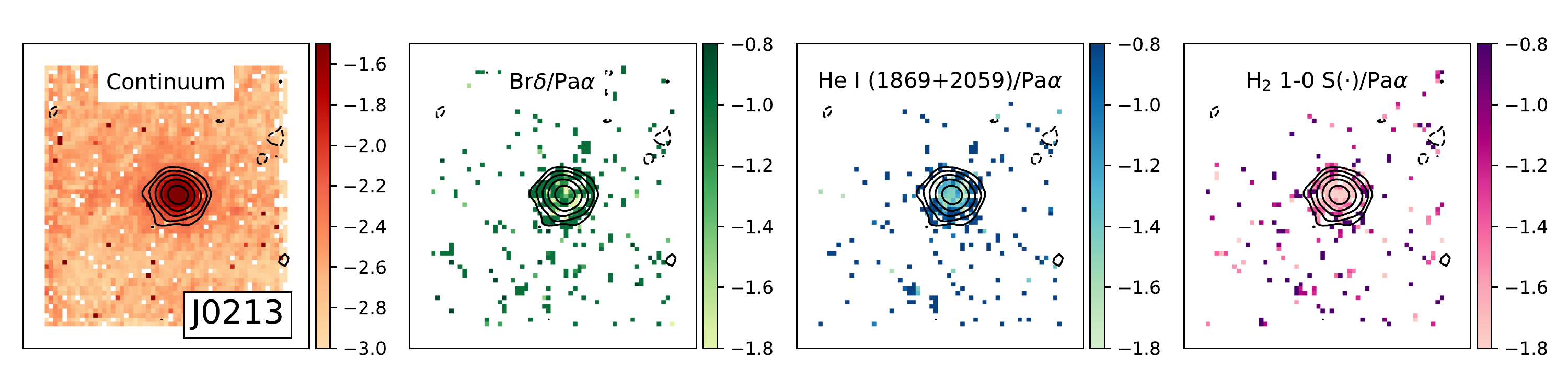}
	\caption{\label{fig:line-ratios}
		     (Continued on next page)}
\end{figure*}
\begin{figure*}
	\includegraphics[width=\textwidth]{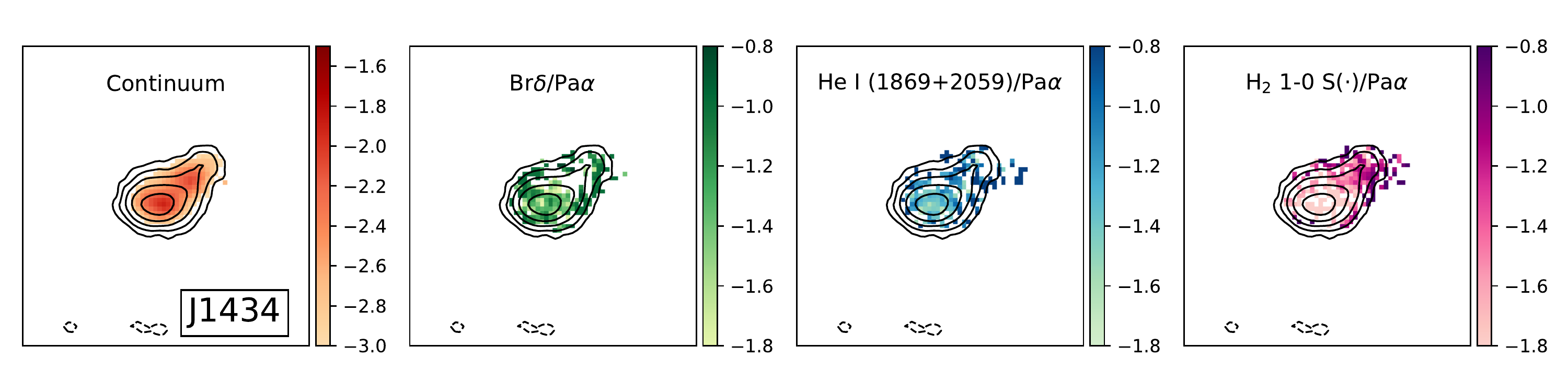}
	\includegraphics[width=\textwidth]{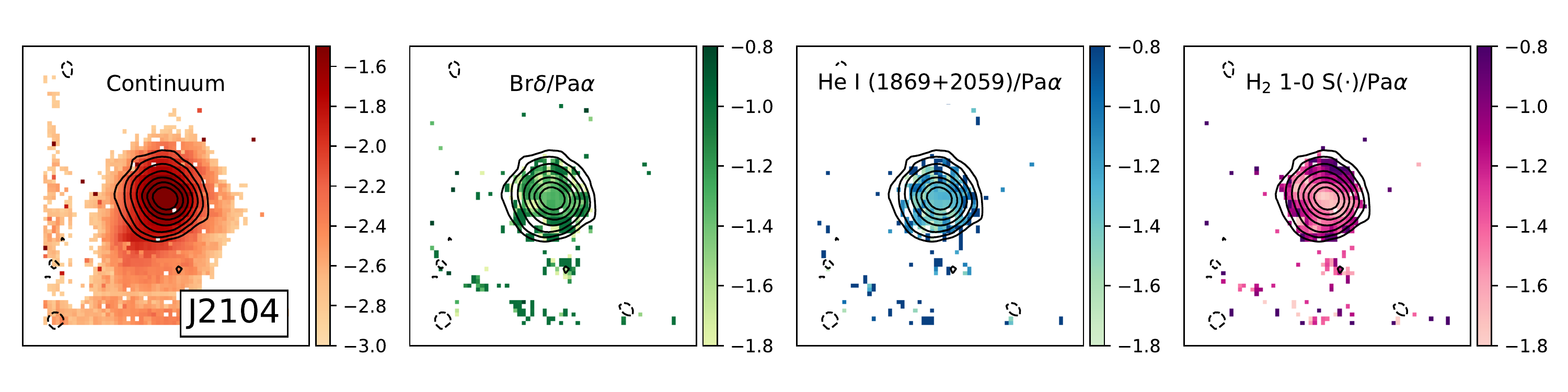}
	\includegraphics[width=\textwidth]{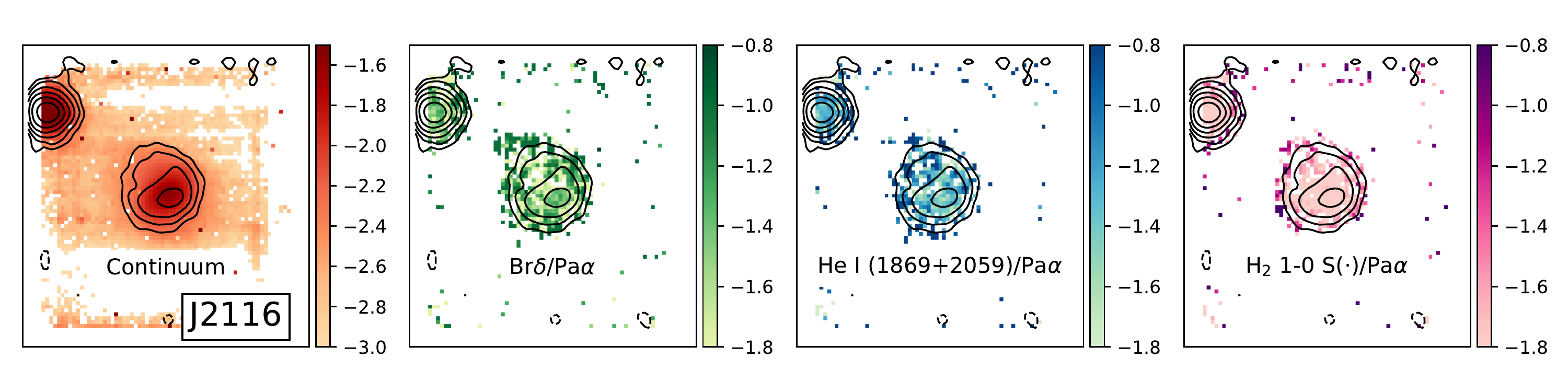}
	\addtocounter{figure}{-1}
	\caption{Continuum and line ratio maps for our LBA sample. 
		Smoothed contours of the jointly-fit Pa$\alpha$ flux are plotted at $\pm 2, 4, 8, \ldots \times \sigma_{\rm Pa\alpha,rms}$.
		Color scales show continuum flux or ratios of the indicated lines in logarithmic units.
		\ion{He}{1} lines are fit together using the same line widths and with 1.869, 1.870, and 2.059~\micron{} line ratios given by \cite{1999ApJ...514..307B} Case B calculations.
		The molecular hydrogen $1-0$~S(1), S(2), S(3), and S(5) lines are jointly fit using the same line widths and with fluxes determined from degeneracies and (unphysical) $T_{\rm ex}= \infty$, which was chosen to maximize signal-to-noise ratio.
		These maps have not been corrected for dust attenuation.
		}
\end{figure*}

We jointly fit the continuum and all recombination and ro-vibrational lines in the NIR spectra using the same methodology as in Section~\ref{sec:results}, except now at \textit{every pixel} and with only a single Pa$\alpha$ Gaussian component.\footnote{
	We attempted to fit the broad Gaussian components as well, but many pixels have low SNR for the Pa$\alpha$ line, which then biases any line ratios that depend on it.
	Thus, we continue using only a single-component Pa$\alpha$ fit.
	We also attempted to fit spectra extracted from $2\times 2$ and $3\times 3$ pixel apertures, but the results were no better than single-pixel spectral fitting.}
From Figure~\ref{fig:line-ratios}, we can compare spatial distributions of continuum, Br$\delta$, \ion{He}{1}, and H$_2$ line ratios relative to Pa$\alpha$.
Pa$\alpha$ moment map contours are also included for visual comparison.
We have masked the edge pixels in all maps, and also applied a $2~\sigma$ cut in Pa$\alpha$ flux to all line ratio maps to prevent noisy pixels from cluttering the plots.

For \ion{H}{1} recombination, we choose to show Br$\delta$ rather than another Brackett series line because it generally has the highest SNR and best-determined fit parameters (see Figure~\ref{fig:recombination-line-fits}).
For the \ion{He}{1}/Pa$\alpha$ ratio, we fit lines using a fixed \ion{He}{1}~1869/\ion{He}{1}~2059 ratio assuming Case B recombination. 
For H$_2$, we fit 1-0 S(1), S(2), S(3), and S(5) ro-vibrational lines assuming a $T_{\rm ex}\rightarrow \infty$ thermally populated spectrum (i.e., we are effectively averaging all of the lines using a 3:1 ortho-to-para ratio in order to maximize the SNR), and plot a single H$_2$~1-0~S($\cdot$)/Pa$\alpha$ ratio.

\subsection{Constraining excitation mechanisms using Pa$\alpha$, H$_2$, and [\ion{Fe}{2}] emission lines} \label{sec:excitation}

\cite{2012A&A...546A..64P} measure resolved Br$\gamma$, H$_2$, and [\ion{Fe}{2}]~1.644~\micron{} emission for a sample of nearby LIRGs and ULIRGs.
If we convert Br$\gamma$ to Pa$\alpha$,\footnote{We assume Case B recombination in order to convert between Br$\gamma$ and Pa$\alpha$ ($T_e=10^4$ K and $n_e = 10^3~{\rm cm}^{-3}$; \citealt{1987MNRAS.224..801H}). 
	We also convert [\ion{Fe}{2}]~1.644~\micron{} to [\ion{Fe}{2}]~1.810~\micron{} using a constant branching ratio of 0.205 (which has no temperature dependence because both transitions have a common upper state; \citealt{1999ApJS..120..101V,2017RMxAA..53..385F}).
} their average flux ratios correspond to log(H$_2$/Pa$\alpha$) $= -1.5$ for the LIRG sample, and log(H$_2$/Pa$\alpha$) $= -2.1$ for the ULIRGs. 
For a subsample of these galaxies, \cite{2015A&A...578A..48C} are able to cleanly separate AGN, supernovae, and young stars as the dominant excitation mechanisms in their sample of (U)LIRGs by comparing resolved [\ion{Fe}{2}]/Br$\gamma$ and H$_2$/Br$\gamma$ line ratios.
Emission lines associated with massive, young stars uniquely populate a locus of low H$_2$/Br$\gamma$ and low [\ion{Fe}{2}]/Br$\gamma$ line ratios, corresponding to $-2.3 \leq \log({\rm H_2 / Pa\alpha}) \leq -1.2$ and $-2.2 \leq $ log([\ion{Fe}{2}]~1.810~\micron{}/$\rm Pa\alpha) \leq -1.4$.

For our LBA sample, [\ion{Fe}{2}]~1.810~\micron{} emission is faint and generally not detected aside from near the Pa$\alpha$ peaks, so we do not show it in Figure~\ref{fig:line-ratios}.
We measure positive [\ion{Fe}{2}]~1.810~\micron{} emission at $>1.5\,\sigma$ significance for J0150, J0213, J2104, and J2116b.
For the central regions of these systems, we find log([\ion{Fe}{2}]~1.810~\micron{}/Pa$\alpha$) $\sim -1.7$.
Despite the large uncertainties, we find that [\ion{Fe}{2}] is too low relative to Pa$\alpha$ for supernovae or AGN to be its primary excitation mechanism \citep[e.g., using the line ratios described by][which are also similar to the results from \citealt{2013MNRAS.429.2587R}]{2015A&A...578A..48C}.
Our well-detected H$_2$/Pa$\alpha$ line ratios reinforce this conclusion: overall, we find that $\log(\rm H_2/Pa\alpha) \lesssim -1.5$, which is in the locus for young, star-forming regions.
LBAs are also similar to other nearby star-forming galaxies in terms of H$_2$/Pa$\alpha$ \citep[albeit with lower ratios than for nearby starbursts with shock excitation; e.g.,][]{1990ApJ...364...77P}.
Therefore, ISM excitation/heating in our sample of LBAs is most impacted by feedback from massive, recently formed stars.

\subsection{Radial gradients in line ratios}
For J2104 (and to an extent, the other LBAs), H$_2$/Pa$\alpha$ line ratios are depressed toward the Pa$\alpha$ emission peak, relative to their values farther from the peak.
We would naively expect Pa$\alpha$ and H$_2$/Pa$\alpha$ to be anticorrelated, but there is an additional more subtle effect at play here.
Specifically, we have only fit single Gaussian profiles to both lines, which may adequately describe the fainter H$_2$ emission (whose broad wings are low in flux, and therefore do not matter much to the model) while failing to fit the brighter Pa$\alpha$ line (whose broad wings force strong constraints on the Gaussian model parameters), such that the single-component Pa$\alpha$ flux is overestimated.
Precisely for this reason, we have used double Gaussian models to measure the total Pa$\alpha$ flux in our previous analysis (e.g., in Section~\ref{sec:recombination}).
We also considered fitting the H$_2$ lines pixel-by-pixel using two-component models, but discovered that the other lines are too faint/noisy to be adequately fit with multiple Gaussians.

Given these difficulties, we avoid comparing line flux ratios for the centers of LBAs, where broad ionized gas components contribute the most to Pa$\alpha$ flux.
We thus interpret the line ratios shown in Figure~\ref{fig:line-ratios} conservatively.
For example, there is an apparent dip in H$_2$/Pa$\alpha$ toward the center of J2104, but we attribute this to the broad, Pa$\alpha$-bright wind that we have reported in Section~\ref{sec:recombination}.
We also find that Br$\delta$/Pa$\alpha$ and \ion{He}{1}/Pa$\alpha$ exhibit similar depressions toward the LBAs' central regions, although the effect seems to be less pronounced.

A high H$_2$/Pa$\alpha$ ratio for peripheral regions of J2104 may indicate that heated molecular gas is prevalent in its outskirts.
It is possible that the ISM conditions are similar to those of the spatially offset H$_2$/Pa$\alpha$ peak in J1434, which we discuss in detail in Section~\ref{sec:merger-shocks}.
For the other LBAs, H$_2$ emission is detected throughout the star-forming, Pa$\alpha$-bright nucleus, but we find no gradient in the H$_2$/Pa$\alpha$ line ratio once we account for broad-line ionized gas components.

\subsection{Evidence for merger-driven shocks?} \label{sec:merger-shocks}

\begin{figure*}
	\plotone{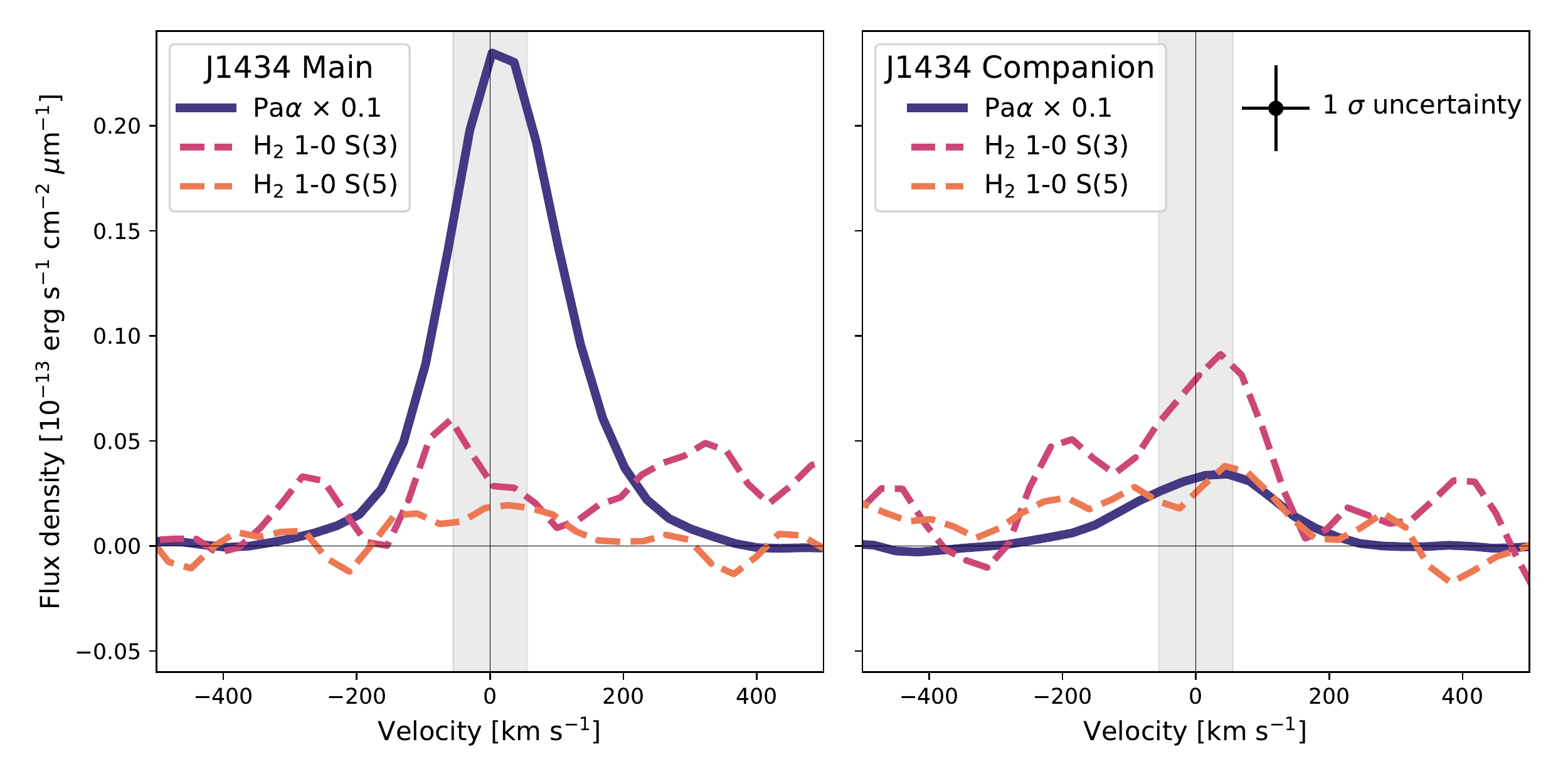}
	\caption{\label{fig:J1434-H_2-line-comparison}
		Pa$\alpha$ (\textit{solid}) and H$_2$ ro-vibrational lines (\textit{dashed}) for J1434.
		We show spectra from the main system (\textit{left}, $z=0.180$) and companion (\textit{right}, $z=0.181$), extracted over two-pixel apertures centered on their respective stellar light peaks and shifted to the same velocity frame.
		The ratios of H$_2$ to Pa$\alpha$ are clearly different for the main source and its companion.
		We have smoothed all spectra using a Gaussian filter with $\sigma=1$~channel for visualization purposes only.
		$1~\sigma$ uncertainties for the H$_2$ lines are shown in the upper right, and the shaded region represents the smoothed spectral resolution.
		Note also that H$_2$~1-0 S(1) does not lie within our spectral range so it is not shown here.
		}
\end{figure*}

The distribution of H$_2$~1-0 ro-vibrational emission in J1434 is concentrated toward the northwest source where the Pa$\alpha$ line is weak.
Near the location of the northwest H$_2$/Pa$\alpha$ peak, we also find a stellar continuum peak.
For the more luminous central Pa$\alpha$ source in J1434, the H$_2$ lines are only weakly detected, implying that its higher rate of star formation may have dissociated H$_2$.
We find $\log(\rm H_2/Pa\alpha) \sim -2$ for the Pa$\alpha$-bright nucleus, and $\log(\rm H_2/Pa\alpha) \sim -1$ for the companion system (which is unusually high).
It is possible that the bright Pa$\alpha$-emitting source is powered by intense star formation activity, while its neighbor to the northwest is affected by AGN, supernovae, or tidal shocks.
Unfortunately, H$_2$/Pa$\alpha$ cannot uniquely identify the dominant sources of this excitation.
\cite{2015A&A...578A..48C} use [\ion{Fe}{2}] emission to break this degeneracy for their sample of local LIRGs and Seyferts, but we are unable to detect the [\ion{Fe}{2}] line at high significance for either component in J1434. 
We place $3~\sigma$ upper limits of log([\ion{Fe}{2}]~1.810~\micron/Pa$\alpha) < -1.4$ for the main system, and $< -0.8$ for the companion.
These line ratios imply that the Pa$\alpha$-bright system is mostly excited by young stars; for the H$_2$-bright companion, low-SNR line ratios are consistent with all excitation mechanisms.
We propose that the enhanced H$_2$ emission in the companion is triggered by merger-driven tidal shocks.

We fit Gaussians to the Pa$\alpha$ and H$_2$ lines for both spatial components, and show the smoothed profiles in Figure~\ref{fig:J1434-H_2-line-comparison}.
The H$_2$ lines in both locations have comparable excitation temperatures consistent with $2000~$K.
We find that, although the H$_2$~1-0 S(3) flux toward the companion is higher ($(4.5 \pm 1.5) \rm \times 10^{-17}~erg~s^{-1}~cm^{-2}~\micron{}^{-1}$) than toward the main system ($(2.8 \pm 1.7) \rm \times 10^{-17}~erg~s^{-1}~cm^{-2}~\micron{}^{-1}$), the Pa$\alpha$ emission from the northwest companion is lower.

\section{Similarities between LBAs and high-redshift galaxies} \label{sec:LBGs-LBAs}

LBAs have moderately strong ionizing spectra, such that they resemble typical $z \sim 2$ star-forming galaxies.
Our sample of LBAs are somewhat dustier than those in previous studies, but are otherwise similar.
We find that they can drive outflows/winds, implying that the neutral gas is somewhat porous even when dust covers the nebular regions \citep[e.g.,][]{2014Sci...346..216B}.
At least one LBA with strong Pa$\alpha$ and \ion{He}{1} emission is marked by high blackbody effective temperatures ($T_{\rm eff} > 40,000$~K).
One possible origin of hot stellar populations is an excess of massive star binaries, which would be consistent with enhanced nebular excitation in high-redshift galaxies based on optical spectroscopy (\citealt{2016ApJ...826..159S}; \citealt{2017ApJ...836..164S}) and X-ray stacks \citep{2013ApJ...762...45B}.
AGN activity is another possible contributor to the high $T_{\rm eff}$, particularly for the higher-mass LBAs that host DCOs, although its effects are subdominant to star formation feedback \citep{2009ApJ...706..203O}.
In general, LBAs are classified as star-forming or BPT composite systems, and the presence of AGN can sometimes be verified by very long baseline radio interferometry.
However, no compact radio continuum emission is detected in the LBA with the most extreme ionization properties \citep[J2104;][]{2012MNRAS.423.1325A}, leading us to surmise that its X-ray emission may be due to a combination of weak/obscured nuclear activity and high-mass X-ray binaries \citep{2011ApJ...731...55J}.
Overall, there is considerable diversity in the correlation between compact radio emission, X-ray emission, excitation properties, and presence of a DCO in LBAs: J0150 and J1434 are detected in X-rays (and J0150 in radio) but have ionization properties otherwise typical of low-$z$ star-forming galaxies, J0213 has a DCO but does not show signs of enhanced excitation, and J2104 has no radio emission but hosts a DCO and is characterized by anomalously high \ion{He}{1}/\ion{H}{1} ratio.

LBAs also have significant warm molecular gas reservoirs that are heated to $\sim 2,000$~K in dense PDRs \citep[rather than shocks; e.g.,][]{Contursi+17}.
NIR line ratios are consistent with feedback from massive stars as the primary excitation source, which supports the interpretation that warm H$_2$ in LBAs traces PDRs nearby star-forming regions.
In one system, we observe excess H$_2$ emission in an interacting companion, which may be evidence of tidal shocks.
In general, LBAs should be thought of as star-forming galaxies that are assembling disks (evidenced by kinematics), and subject to strong but not catastrophic feedback, similar to distant LBGs and other typical high-$z$ star-forming galaxies in the same mass range. 
At least some of the detailed differences between objects in our LBA sample (see above) may reflect diverse recent histories of dynamical interactions.

\section{Conclusions} \label{sec:conclusions}

We have obtained $\sim 0.7\arcsec$ resolution integral field spectroscopy for a sample of six UV-bright galaxies at $z = 0.1-0.3$ selected from the \cite{2005ApJ...619L..35H} SDSS sample, which resemble high-redshift Lyman break galaxies in terms of SFRs, stellar masses, metallicities, and rest-frame UV sizes.
Physical conditions in their ISMs are comparable to those of typical $z\sim 1-2$ star-forming galaxies.
We therefore consider our sample to be low-redshift Lyman break galaxy analogs (LBAs).
One LBA system, J2116, consists of two bright sources with rotating gas disks.

LBAs also share similarities to high-$z$ LBGs and star-forming galaxies in terms of their kinematics and dust attenuation properties.
By analyzing their hydrogen, helium, [\ion{Fe}{2}], and H$_2$ ro-vibrational emission lines, we are able to probe the physical processes that heat and excite the multiphase gas in LBAs.
For our sample of relatively massive and dusty LBAs, star formation is primarily responsible for exciting warm H$_2$ in dense PDRs, although feedback from obscured AGN and merger-driven shocks may be subdominant contributors to ISM heating in some systems. 
These feedback processes are capable of producing outflows and large gas velocity dispersions.
Overall, we find that LBAs exhibit strong but not overwhelming feedback, similar to typical star-forming galaxies at high redshifts.

We list our detailed conclusions below:
\begin{enumerate}
	\item Using the Pa$\alpha$ line, we have traced the kinematics of the ionized gas and produced moment maps. 
	We measure inclination-corrected ordered velocity-to-dispersion ratios $\langle v/\sigma \rangle \approx 1.8-2.2$ from moment maps, and $\sim 1.2$ by kinematically modeling the $3d$ data cube. 
	We find values that are much lower than typical $v/\sigma$ for low-$z$ star-forming spirals, similar to other LBA \citep{2010ApJ...724.1373G} and LBG \citep{2015ApJ...799..209W} samples.
	Mean velocity dispersions in the LBA sample are very high: $\langle \sigma \rangle = 75-90~\rm km~s^{-1}$, depending on how they are measured.
	\item We use H$\alpha$, H$\beta$, and Pa$\alpha$ line ratios to measure the nebular color excess under the assumption of a Calzetti attenuation curve.
	For our sample, we derived nebular color excesses $\langle E(B-V)_{\rm neb}\rangle = 0.26 \pm 0.04$, $0.32 \pm 0.11$, and $0.35 \pm 0.17$, using the H$\alpha$ + H$\beta$, Pa$\alpha$ + H$\beta$, and Pa$\alpha$ + H$\alpha$ line ratios, respectively.
	The average continuum-to-nebular color excess ratio changes as a function of the optical depth (or recombination line) used to probe \ion{H}{2} regions.
	The observed attenuation is consistent with a well-mixed model of ionizing and non-ionizing stars (e.g., closer to a single effective screen) when we use only the bluer Balmer lines.
	The color excess ratio probed by Pa$\alpha$ is more similar to a lower, Calzetti-like ratio of continuum-to-nebular extinction, $f = 0.44$, indicating an extra layer of attenuation toward nebular regions.
	\item We detect Brackett series hydrogen recombination and \ion{He}{1} helium recombination lines at 1.869+1.870~\micron{} (blended) and 2.059~\micron{}.
	We model the effective blackbody temperature of an ensemble of young stars and the ionization parameter $\log U$ and compare with ratios of the Pa$\alpha$, Br$\delta$, \ion{He}{1}~1869+1870, and \ion{He}{1}~2059 lines.
	We find that most objects in our sample are consistent with $32,000~{\rm K} < T_{\rm eff} < 40,000~{\rm K}$ and $\log U = -3$, with the three strongest Pa$\alpha$ emitters also characterized by the highest average effective blackbody temperatures. 
	J2104 appears to require a higher $T_{\rm eff}$, which, in conjunction with its broad Pa$\alpha$ line width, may be evidence of an AGN, a top-heavy IMF, or the presence of massive (X-ray) binaries.
	\item We measure dust-corrected H$_2$~1-0~S(1) through S(7) ro-vibrational line fluxes and widths.
	We find $T_{\rm ex} \sim 2000$~K single-temperature H$_2$ gas reservoirs, non-detection of H$_2$~2-1~S$(\cdot)$ lines, and faint [\ion{Fe}{2}] emission lines in some cases.	
	Our new results reinforce previous evidence \citep[e.g.,][]{Contursi+17} that H$_2$ ro-vibrational emission lines in LBAs originate from UV- and/or X~ray-irradiated dense PDRs rather than shocks.
	\item We use H$_2$ line fluxes to estimate the mass of the warm molecular ISM.
	The global warm molecular masses range from $(1.8 - 38) \times 10^{3}~M_\sun$.
	Warm-to-cool H$_2$ mass ratios range between $(6.2 - 15.1) \times 10^{-7}$.
	When compared with nearby prototypes of a galaxy-wide starburst, Type~1 quasar, and Type~2 Seyfert, the LBAs most closely resemble the third, indicating that their H$_2$ properties may be similar to those of an obscured AGN.
	\item We fit the spectra at every pixel for each LBA, allowing for investigation of the spatially resolved distributions of stellar continuum, ionized gas (using helium and multiple hydrogen lines), and warm molecular gas.
	We measure log(H$_2$/Pa$\alpha) \lesssim -1.5$ for most regions in the sample, and faint [\ion{Fe}{2}] emission, which are consistent with massive, recently formed stars as the primary sources of excitation for the nebular gas.
	The observed line ratios rule out AGN or supernovae as dominant sources of feedback in the centers of LBAs. 
	For J1434, H$_2$/Pa$\alpha$ in its northwestern companion source is anomalously strong, which can be interpreted as evidence for merger-driven shocks.
\end{enumerate}

These results blaze a trail for future rest-NIR observations of $z \sim 3$ LBGs in the {\it James Webb Space Telescope} era, which will allow us to assess how far we can extend the analogy with LBAs in terms of the latter population's diverse but overlapping properties.

\bibliographystyle{apj}
\bibliography{bibliography.bib}

\acknowledgments

The authors thank Natascha F\"orster Schreiber and Sabine Mengel for their help preparing the SINFONI observations and the staff of the ESO Paranal Observatory for carrying them out.
The authors also thank Whitney Kropat and Kelsey Funkhouser for their contributions to the early stages of this analysis, and Jane Rigby and Amiel Sternberg for useful discussions.
The authors additionally thank the anonymous referee for providing useful comments and suggestions that have improved this manuscript.
JFW and AJB acknowledge significant support for this work from the U.S. National Science Foundation through grant AST-0955810.

Funding for the Sloan Digital Sky Survey IV has been provided by the Alfred P. Sloan Foundation, the U.S. Department of Energy Office of Science, and the Participating Institutions. SDSS-IV acknowledges
support and resources from the Center for High-Performance Computing at
the University of Utah. The SDSS web site is www.sdss.org.

SDSS-IV is managed by the Astrophysical Research Consortium for the 
Participating Institutions of the SDSS Collaboration including the 
Brazilian Participation Group, the Carnegie Institution for Science, 
Carnegie Mellon University, the Chilean Participation Group, the French Participation Group, Harvard-Smithsonian Center for Astrophysics, 
Instituto de Astrof\'isica de Canarias, The Johns Hopkins University, Kavli Institute for the Physics and Mathematics of the Universe (IPMU) / 
University of Tokyo, the Korean Participation Group, Lawrence Berkeley National Laboratory, 
Leibniz Institut f\"ur Astrophysik Potsdam (AIP),  
Max-Planck-Institut f\"ur Astronomie (MPIA Heidelberg), 
Max-Planck-Institut f\"ur Astrophysik (MPA Garching), 
Max-Planck-Institut f\"ur Extraterrestrische Physik (MPE), 
National Astronomical Observatories of China, New Mexico State University, 
New York University, University of Notre Dame, 
Observat\'ario Nacional / MCTI, The Ohio State University, 
Pennsylvania State University, Shanghai Astronomical Observatory, 
United Kingdom Participation Group,
Universidad Nacional Aut\'onoma de M\'exico, University of Arizona, 
University of Colorado Boulder, University of Oxford, University of Portsmouth, 
University of Utah, University of Virginia, University of Washington, University of Wisconsin, 
Vanderbilt University, and Yale University.

\end{document}